\documentclass[twocolumn,prd,aps,superscriptaddress,preprintnumbers,tightenlines,showpacs,nofootinbib,eqsecnum,amsfonts,amsmath]{revtex4}
\pdfoutput=1

\usepackage{color}
\usepackage{calc}
\usepackage{amsmath,amssymb,graphicx}
\usepackage{tensor}
\usepackage{bm}
\usepackage{times}
\usepackage[varg]{txfonts}
\usepackage{float}
\usepackage{dcolumn}

\usepackage[nolist,nohyperlinks]{acronym}
\usepackage{xspace}

\usepackage[abs]{overpic}
\usepackage{pict2e}

\usepackage[caption=false]{subfig} 

\usepackage[utf8]{inputenc}

\usepackage{hyperref} 
\usepackage{braket}
\usepackage{pgffor}
\allowdisplaybreaks
\usepackage{float}
\usepackage{listings}
\usepackage{tensor}
\usepackage{ulem}
\usepackage[makeroom]{cancel}
\usepackage{orcidlink}

\newcommand{\ICS}{Astrophysik-Institut, Universit\"{a}t Z\"{u}rich, Winterthurerstrasse 190, 8057 Z\"{u}rich, Switzerland}
\newcommand{\UZH}{Physik-Institut, Universit\"{a}t Z\"{u}rich, Winterthurerstrasse 190, 8057 Z\"{u}rich, Switzerland}
\newcommand{\APC}{Université Paris Cité, CNRS, Astroparticule et Cosmologie, F-75013 Paris, France}

\begin{document}
\title{GPU-accelerated LISA parameter estimation with full time domain response}

\author{Cecilio García-Quirós\, \orcidlink{0000-0002-8059-2477}}
\email[Correspondence email address: ]{cecilio.garciaquiros@uzh.ch}
\affiliation{\ICS}\affiliation{\UZH}\affiliation{\APC}
    
\author{Shubhanshu Tiwari\,\orcidlink{0000-0003-1611-6625}}
\email{shubhanshu.tiwari@physik.uzh.ch}
\affiliation{\UZH}

\author{Stanislav Babak\, \orcidlink{0000-0001-7469-4250}}
\email{stas@apc.in2p3.fr}
\affiliation{\APC}

\date{\today} 

\begin{abstract}
We conduct the first full Bayesian inference analysis for LISA parameter estimation incorporating the effects of subdominant harmonics and spin-precession through a full time domain response. The substantial computational demands of using time domain waveforms for LISA are significantly mitigated by a novel Python implementation of the \textsc{IMRPhenomT} family of waveform models and the LISA response with GPU acceleration. This time domain response alleviates the theoretical necessity of developing specific \textit{transfer functions} to approximate the LISA response in the Fourier domain for each specific type of systems and allows for the use of unequal-arms configurations and realistic LISA orbits.
Our analysis includes a series of zero-noise injections for a Massive Black Hole Binary with aligned and precessing spins. We investigate the impact of including subdominant harmonics, compare equal and unequal-arm configurations, and analyze different Time-Delay-Interferometry (TDI) configurations. We utilize full and uniform priors, with a lower frequency cutoff of 0.1 mHz, and a signal duration of approximately two months, sampled every 5 seconds. The sampler is initialized based on Fisher's estimates. Our results demonstrate LISA's capability to measure the two spin magnitudes and the primary spin tilt angle, alongside sky localization, with percent-level precision, while component masses are determined with sub-percent accuracy.
\end{abstract}


\maketitle

\section{Introduction} \label{sec:introduction}
The Laser Interferometer Space Antenna (LISA) mission \cite{lisaredbook} was officially adopted by ESA and NASA in January 2024, with a scheduled launch for 2035. LISA will deploy three identical spacecraft into space to form the first space-based gravitational wave (GW) detector. Operating in space allows LISA to bypass many of the noise limitations that terrestrial detectors face \cite{Abbott_2020}, enabling the observation of a new class of GW sources. LISA will observe in the millihertz frequency range, detecting signals from a broad spectrum of sources, from stellar-mass compact objects to supermassive black holes, providing us with unprecedented insights into fundamental physics. LISA promises to be a transformative mission, set to revolutionize our understanding of the universe and GW astronomy.


The LISA mission presents an extraordinary opportunity for scientific advancement but also introduces significant challenges both on the technological and data analysis fronts. To maximize the extraction of physical information from LISA's observations, these challenges must be addressed effectively. LISA will operate at much lower frequencies than current ground-based detectors, allowing for the detection of much longer gravitational wave signals. Coupled with a vast increase in the number of events, this necessitates the development of waveform models that are computationally efficient to ensure parameter estimation remains feasible (see \cite{lisawaveforms} for a review of the challenges that waveform models will face for LISA).

Moreover, LISA's ability to observe a wide range of masses requires waveform models that encompass an expanded parameter space. At the same time, the high signal-to-noise ratios expected for LISA events will demand highly accurate models capable of capturing subtle physical effects that are less noticeable with current detectors. Key among these effects are precession and eccentricity, which provide crucial information about the astrophysical formation and evolutionary pathways of compact binaries \cite{Breivik_2016, Steinle_2021,  Zevin_2021, Fumagalli_2023, Fumagalli_2024}. Developing efficient parameter estimation techniques that incorporate both precession and eccentricity will be of paramount importance for fully exploiting LISA’s scientific potential.

Parameter estimation for gravitational waves is typically performed in the Fourier domain due to the easier characterization of the noise and the efficiency and sensitivity of the \textit{matched filtering} technique \cite{Bruce_2003, Bruce_2012}, where the complicated convolutions and correlations between templates and data become a simple multiplication in the Fourier domain. Consequently, waveform models formulated directly in the Fourier domain \cite{phenomx, phenomxhm, phenomxphm, Cotesta_2020} have gained significant popularity in GW data analysis \cite{gwtc3, Marta_2021, Maite_2022}. By avoiding the need for a discrete Fourier transform of each template, these models save valuable computational resources. In contrast, time domain waveforms require fine sampling on a uniformly spaced time grid to facilitate the application of the Fast Fourier Transform (FFT) algorithm. This constraint makes time domain waveforms particularly inefficient when dealing with long-duration signals, as the sampling grid must encompass the entire signal duration, even if one is interested only in a reduced frequency range. This also prevents the use of custom nonuniform frequency grids, which could be used to optimize performance \cite{Vinciguerra_2017, Multibanding_2021, ROQs}.


However, from the modelling perspective, the time domain is often preferred because it naturally represents the temporal evolution of physical systems. Numerical Relativity (NR), Effective One Body formalism (EOB) \cite{eob_99, eob_09} or post-Newtonian (PN) theory \cite{Clifford_2011, Blanchet_2014} solve the dynamic equations of motion in the time domain, yielding time domain waveforms that accurately describe the intricate behaviour of binary systems. In this domain, it is easier to incorporate physical processes and time-varying interactions, making it easier to asses how physical details like spin precession or eccentricity affect the waveform's shape. Moreover, certain phenomena like displacement memory, a permanent change in spacetime following the passage of gravitational waves, are more naturally described and understood in the time domain \cite{Maria}. As LISA aims to explore new physics through these effects, time domain models will be crucial for maximizing the scientific return of the mission, enabling precise waveform modelling that is harder to replicate when relying solely on frequency domain representations.

Apart from the challenges in waveform modelling, LISA introduces the additional complexity of computing a dynamic detector response. For current ground-based detectors, signals are typically short-lived, allowing the detector to be approximated as static. In these cases, the detector response is simply a linear combination of the two gravitational wave polarizations, with the response factors depending only on the sky position and polarization angle, but not on time.
However, LISA’s sources can remain in-band for extended durations, from days to months or even years \cite{lisaredbook}, causing the signal to be modulated by the orbital motion of the LISA constellation. Moreover, the laser frequency noise is several orders of magnitude larger than the GW signal LISA aims to detect. This noise is exacerbated when the LISA arms have unequal length. In a realistic scenario, the LISA spacecraft will not maintain a perfect equilateral triangular formation, leading to time-varying arm lengths and complicating gravitational wave detection.
Fortunately, the Time-Delay Interferometry (TDI) technique \cite{tdi} mitigates this issue by combining signals from different arms in a carefully timed manner, effectively cancelling out laser frequency noise by several orders of magnitude. TDI channels reconstruct gravitational wave signals while suppressing noise, thereby enhancing the sensitivity of the detector to gravitational waves.

The LISA arm response and TDI are better understood and described in the time domain. In this domain, they can be applied to the GW $h_{+,\times}(t)$ polarizations exhibiting any physical effect: higher harmonics, precession, eccentricity, or other complex features. However, applying these operations in the Fourier domain is more challenging and requires the development of theoretical \textit{transfer functions} \cite{Marsat_2018,Marsat_2021} to approximate the LISA response. These transfer functions rely on the assumption that a unique time-frequency correspondence exists in the gravitational wave signal. The requirement for a monotonic frequency and amplitude evolution complicates the Fourier domain approach, particularly for complex signals. Marsat et al. \cite{Marsat_2021} addressed this issue for subdominant harmonics signals in aligned spin systems by developing a specific transfer function for each harmonic. Later, Pratten et al. \cite{Geraint_2023} extended it to precessing systems by introducing the multiple-scale analysis to provide first-order corrections to the transfer function for each harmonic in precessing systems. These works marked significant progress, as they were the first to perform full Bayesian inference using complete inspiral-merger-ringdown (IMR) Fourier-domain waveforms for LISA. In contrast, Garg et al. \cite{Garg_2024} applied this formalism to eccentric systems, using a Fourier-domain inspiral-only approximant without higher harmonics. So far, there exists no transfer function for the GW memory. These developments illustrate the evolving complexity of accurately modelling the LISA detector response in the frequency domain, particularly for complex signals such as those involving precession, eccentricity or GW memory. Furthermore, all these works assume an equilateral configuration of the LISA constellation.

While the transfer function formalism shows promising results, it relies on several assumptions, like the equal-arms configuration, which need to be validated against a more general formalism. The time domain approach can incorporate more realistic LISA configurations, providing a means to evaluate the robustness of the Fourier domain method. In this study, we take the first step in this direction by enabling the possibility of performing full LISA parameter estimation (PE) using time domain waveforms and responses. This is achieved by substantially reducing the computational cost of the likelihood evaluations for time domain signals.

To achieve this, we will exploit the parallelization capabilities of phenomenological waveform models and employ a new implementation of the \textsc{IMRPhenomT} family \cite{phenomt, phenomthm, phenomtphm} and the LISA response with GPU support. We will conduct a series of signal injections, including aligned-spin and precessing systems with subdominant harmonics, to examine the accuracy with which LISA can recover source parameters. Additionally, we will investigate the influence of subdominant harmonics in the waveforms, the impact of unequal-arms orbits, the consequences of omitting certain TDI channels and the use of different TDI generations. These studies will provide key insights into the accuracy and efficiency of LISA parameter estimation methods.


\section{Methods} \label{sec:develop}

\subsection{Waveform generation}
The waveform generation, i.e. the calculation of the polarizations $h_{+,\times}(t)$, will be executed using a newly developed implementation of the \textsc{IMRPhenomT} family within the Python programming language with GPU support. \textsc{Phenom} models often utilize analytical expressions to describe the waveform as functions of time or frequency, where each time/frequency data point is computed independently from the rest. This characteristic permits the parallelization over the time/frequency array, a feature incorporated into this implementation through the use of Python libraries like \texttt{numpy} \cite{numpy} and \texttt{numba} \cite{numba} for CPU operations, as well as \texttt{cupy} \cite{cupy} for GPU acceleration. Importantly, this new implementation has been designed with agnostic code, allowing it to operate effectively on both CPU and GPU architectures. \texttt{BBHx} \cite{bbx1, bbx2} is a similar software package designed for GPU implementation, supporting earlier generations of Fourier-based \textsc{Phenom} models, such as \textsc{PhenomHM} \cite{phenomhm}, in addition to the Fourier LISA response.

The new implementation is designed to replicate the functionality found in the LIGO Algorithms Library (LALSuite) \cite{lalsuite, swiglal}, but has been specifically tailored to extend the modularity and usability of phenomenological waveform models beyond the LIGO-Virgo-Kagra (LVK) infrastructure and its conventional sources. By facilitating a Python-based framework, this new package alleviates the challenges faced by waveform developers who traditionally need to reimplement their models in C99, the language supported by LALSuite. The adoption of Python will allow for easier integration of advanced modeling techniques, such as machine learning and neural networks, which benefit from the more extensive library support available in Python.

Moreover, the \textsc{Phenom} modelling strategy also facilitates the implementation of advanced algorithms that leverage nonuniform time and frequency grids, significantly reducing the number of required points. Notable methods include Multibanding \cite{Vinciguerra_2017, Multibanding_2021, Soichiro_2021}, heterodyne likelihood \cite{Cornish_2021}, and relative binning \cite{Zackay_2018, Leslie_2021, Krishna_2023}. Efforts are also underway to adapt these algorithms for use in the time domain, further enhancing their applicability. With this flexible implementation, we anticipate a significant boost in the development and adoption of phenomenological waveform models, which are crucial to meet the computational efficiency and accuracy requirements, not only for LISA, but also for future third-generation ground-based detectors like Einstein Telescope \cite{Maggiore_2020} and Cosmic Explorer \cite{Reitze_2019}. 

The current implementation includes the complete \textsc{IMRPhenomT} family: \textsc{IMRPhenomT} \cite{phenomt} (aligned-spin, (2,2) mode only), \textsc{IMRPhenomTHM} \cite{phenomthm} (aligned-spin with subdominant modes), \textsc{IMRPhenomTP} \cite{phenomtphm} (spin-precession with only the (2,2) mode in the co-precessing frame), \textsc{IMRPhenomTPHM} \cite{phenomtphm} (spin-precession with co-precessing subdominant modes). In the precessing sector, the GPU acceleration is only fully implemented for the next-to-next leading order (NNLO) \cite{Bohé_2013} prescription of the Euler angles describing the precessing motion. The Multi-scale Analysis (MSA) angles \cite{MSA} will be incorporated soon. The numerical angles obtained from the evolution of the SpinTaylor equations \cite{Buonanno_2004} can only use the GPU by interpolating or transfering the full ODE solution to the GPU. Our analysis will only employ the NNLO analytical prescription with GPU support. By using this single-spin approximation, we will miss double-spin information and not measure very well the tilt and azimuthal angle of the secondary spin.

Another notable advancement over the LALSuite implementation involves the precessing sector. The \textsc{Phenom} approach approximates precessional effects using the \textit{twisting-up} technique \cite{Patricia_2012, Patricia_2015}, which establishes a co-precessing frame that dynamically follows the motion of the orbital angular momentum (or the direction of maximum emission). This is achieved through a set of Euler angles, with various prescriptions available, such as the NNLO, MSA or SpinTaylor mentioned above.
In the co-precessing frame, the waveform resembles an aligned spin one and is generated with the aligned spin model. Then it is transformed to an inertial frame by a time-dependent Euler angle rotation which incorporates the waveform modulations proper of precession. In \textsc{IMRPhenomT}, two distinct rotations are employed: the first transforms the waveform from the co-precessing (CP) frame to the intermediate inertial $J$-frame, while the second transforms from the $J$-frame to the final $L_0$-frame which is the reference frame used in the LALSuite conventions. These two rotations and the calculation of the polarizations are formulated as 
\begin{align}
& h^J_{\ell m}(t) = \sum_{m^{\prime} = - \ell}^\ell \mathcal{D}^{\ell \ast}_{m m^{\prime}} \left(\alpha(t), \beta(t), \gamma(t) \right) \, h^{CP}_{\ell m^{\prime}}(t)\\
            & h^{L0}_{\ell m}(t) = \sum_{m^{\prime} = - \ell}^\ell \mathcal{D}^{\ell \ast}_{m m^{\prime}} \, (\alpha_0, \beta_0, \gamma_0) \, h^{J}_{\ell m^{\prime}}(t)\\ 
            & h_+ -i h_\times = \sum_{l=2}^\infty \sum_{m=-l}^{l} h^{L_0}_{lm}(t) \tensor[_{-2}]{Y}{_{lm}}(\iota, \phi),
\end{align}
where $\mathcal{D}^{\ell}_{m m^{\prime}}$ are the Wigner-D matrices and $\alpha$, $\beta$, $\gamma$ are the Euler angles. With this strategy, the five harmonics in the co-precessing frame included in \textsc{IMRPhenomT} require around 250 operations for each time step, including the complex exponentials in the Wigner-D matrices which are computationally expensive. A model with all the $l=5$ modes in the co-precessing frame would instead require 321 and one with all the $l=8$, 1939 operations.

In the new implementation, we utilize the efficient calculation of polarizations in rotating frames detailed in Appendix B of \cite{Boyle}, which we summarize in the following. By leveraging the relationship between the Wigner-D matrices and the spin-2 weighted spherical harmonics \cite{Boyle_2016} one obtains that the polarizations can be written as
\begin{equation}\label{eq:newtwistup}
    h_+ - i h_{\times} = \sum_{l,m} h^{CP}_{lm}(t)\: \sqrt{\frac{2l+1}{4\pi}}\mathcal{D}_{m,2}^l(R_{CP-J}R_{J-L_0}R_{\iota, \phi}).
\end{equation}
Thus, the computational complexity is reduced to a single summation over the number of modes in the co-precessing frame. The number of operations per time step is reduced to 10, 32, and 77 for the \textsc{IMRPhenomT}, a $l=5$ and a $l=8$ co-precessing model respectively. The Wigner-D matrices now encapsulate the resultant combinations of the two rotations between frames $R_{CP-J}$, $R_{J-L_0}$ along with the alignment with the line-of-sight $R_{\iota, \phi}$\footnote{The corresponding Euler angles for this last rotation are $\alpha=\phi$, $\beta=\iota$, $\gamma=0$.}.

Consecutive rotations, as the ones considered here, can also be represented as a unique rotation characterized by a new set of Euler angles. However, calculating these new angles is highly complex. Each time step would require computing the rotation matrices from the Euler angles of each rotation (which involves trigonometric functions), performing the matrix multiplication and extracting the angles through inverse trigonometric functions. Besides being computationally intensive, this method is also very sensitive to numerical errors\footnote{For drawbacks of using Euler angles see \cite{BoyleEuler}.}.

A more efficient approach for handling rotations involves using \textit{quaternions} instead of Euler angles. Quaternions are four-dimensional constructs that obey specific rules of addition and multiplication, often used to represent rotations. The primary advantage is that successive rotations can be combined by simply multiplying quaternions, eliminating the need for complex matrix multiplications or costly inverse trigonometric functions at each time step. In this approach, we convert the Euler angles of each rotation into quaternions, then apply the quaternion multiplication to combine these rotations. The resulting quaternion is then used to compute the Wigner-D matrices in Eq.~\ref{eq:newtwistup}. Given a quaternion $q=(q_0,q_1,q_2,q_3)$, its components are derived from the Euler angles as
\begin{equation}
\begin{aligned}
    q_0 &= \sqrt{1 + \cos \beta} \: \cos \frac{\alpha + \gamma}{2}, \\
    q_1 &= -\sqrt{1 - \cos \beta} \: \sin \frac{\alpha - \gamma}{2}, \\
    q_2 &= \sqrt{1 - \cos \beta} \: \cos \frac{\alpha - \gamma}{2}, \\
    q_3 &= \sqrt{1 + \cos \beta} \: \sin \frac{\alpha + \gamma}{2}, \\
\end{aligned}
\end{equation}
and the Wigner-D matrix as
\begin{equation}
\begin{aligned}
Q_a &= q_0 + i q_3\\
Q_b &= q_2 + i q_1\\
\mathcal{D}_{m,m'}^l(q) &= \sqrt{\frac{(\ell+m)!(\ell-m)!}{(\ell+m')!(\ell-m')!}} | Q_a |^{2\ell-2m} Q_a^{m'+m} Q_b^{-m'+m}\\
& \times \sum_{\rho=0}^{l+m'} \binom{\ell+m'}{\rho} \binom{\ell-m'}{\ell-\rho-m}
\left( - \frac{|Q_b|^2}{|Q_a|^2} \right)^\rho.
\end{aligned}
\end{equation}

This quaternion-based formalism is general and could also be applied in the Fourier domain for the \textsc{IMRPhenomX} family. The potential speed-up may not be as significant as for \textsc{IMRPhenomT}, given that \textsc{IMRPhenomX} involves only one rotation—from the co-precessing frame to the $J$-frame. Additionally, in \textsc{IMRPhenomX}, the Euler angles are computed at rescaled frequencies for each co-precessing mode, requiring the computation of more quaternions than in the time domain approach.

\begin{figure}
    \centering
    \includegraphics[width=\columnwidth]{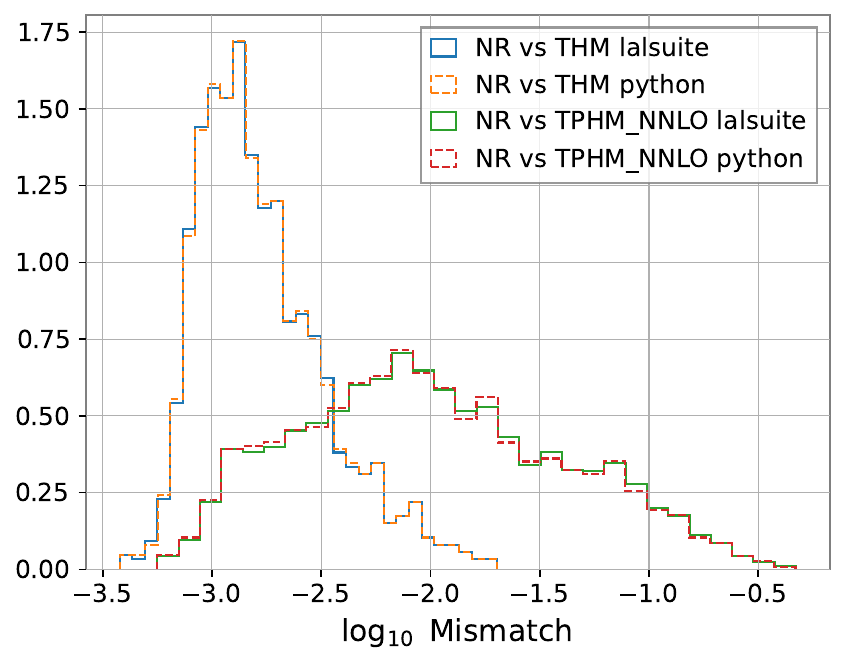}
    \caption{Mismatch distributions for the LALSuite and python implementations against NR SXS simulations. The aligned-spin model (\textsc{IMRPhenomTHM}) is matched against 1506 cases and the precessing version (\textsc{IMRPhenomTPHM\_NNLO}) against 4152 in the mass range $[10^6,10^7]M_\odot$.}
    \label{fig:mismatches}
\end{figure}

The new Python code also refines some aspects of the LALSuite implementation and enhances numerical stability by replacing finite-difference derivatives with analytical expressions, along with other improvements. We assess the accuracy of the new implementation by benchmarking it against numerical relativity waveforms from the \texttt{LVCNR-SXS} catalog~\cite{nrinjectioninfrastructure, Boyle_2019} and compare the results to those obtained with the LALSuite implementation to ensure reliability.
We evaluate 502 aligned-spin and 1384 precessing simulations from the SXS catalog~\cite{Boyle_2019}. For each configuration, we generate three different cases by uniformly sampling the total mass in the range $[10^6, 10^7]\,M_{\odot}$, the inclination angle from $[0, \pi]$ radians, the reference phase $\phi_\mathrm{ref}$ from $[0, 2\pi]$ radians and the luminosity distance from $[1, 10]$~Gpc. The minimum and reference frequencies are taken from the NR simulation metadata. In some cases, the simulations do not extend to sufficiently low frequencies (i.e., $10^{-4}$~Hz); in such cases, we compute the match starting from the minimum frequency available. 
We use a sampling interval of $\Delta t = 5$~s and a frequency resolution of $\Delta f = 10^{-6}$~Hz, corresponding to a time duration of approximately 11.6 days. The match is computed directly on the plus polarization using the sensitivity curve \texttt{analytical\_psd\_lisa\_tdi\_AE} as implemented in the \texttt{PyCBC} package~\cite{pycbc}. The mismatches are numerically optimized over the reference phase $\phi$ for the aligned-spin cases and over the reference phase and solid rotations of spins vectors for the precessing cases.

The results are summarized in Fig.~\ref{fig:mismatches}.
Our aim here is not to carry out a detailed mismatch study or to evaluate whether these models satisfy the LISA accuracy requirements, but rather to compare the mismatch values between the two implementations and verify that no significant differences arise. Consequently, the absolute mismatch values presented here should not be interpreted as statements about the overall waveform accuracy for LISA.

The results indicate that the mismatch distributions are nearly identical compared to NR, suggesting that the small improvements introduced in the Python implementation are negligible relative to the inherent modeling systematics with NR. Similar conclusions (not shown) are obtained by comparing with \textsc{SEOBNRv5PHM}~\cite{seobnrv5phm} waveforms in both the LISA and LVK parameter regimes.
We conclude that the new implementation is robust and suitable for use in this work. However, we stress that we are not claiming that these waveforms meet the full accuracy requirements of LISA. In the following section, we perform injections and recoveries using the same waveform model. This allows us to assess how well parameters could be recovered under the assumption of a perfect waveform model, with the understanding that this idealization does not hold in practice. The primary objective here is to demonstrate that parameter estimation using the full time-domain LISA response is computationally feasible, even for precessing binary systems.

\begin{figure}
    \centering
    \includegraphics[width=\columnwidth]{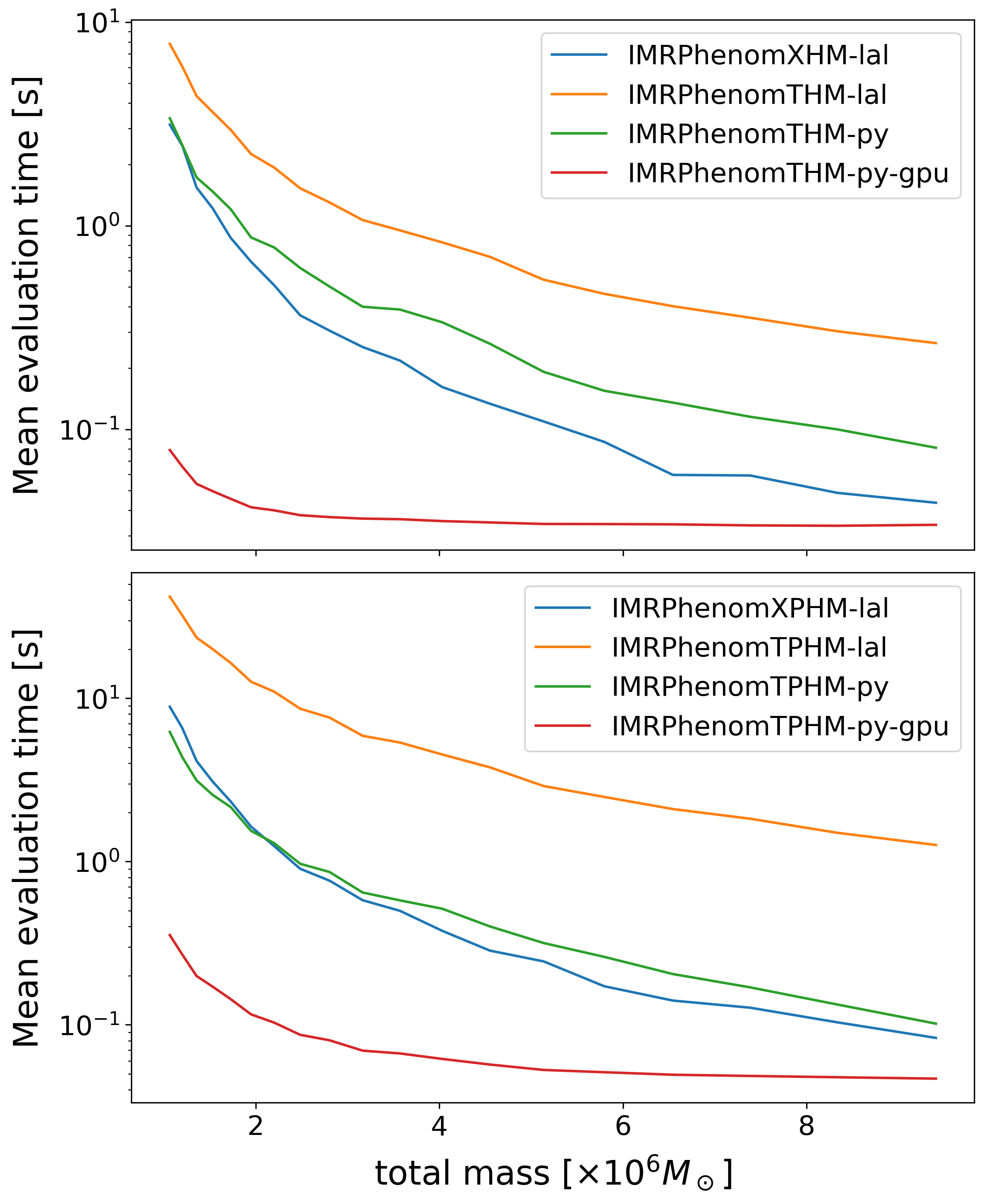}
    \caption{Mean evaluation time of a waveform generation (in Fourier domain) for different models and implementations as a function of total mass. Top panel shows the aligned spin models while the bottom compares the precessing ones. The average is taken over 1000 randomly distributed cases with a frequency range of 0.1mHz-0.1Hz, therefore with a time sampling of 5s.}
    \label{fig:timings}
\end{figure}

In Fig.~\ref{fig:timings}, we show the average computational cost of a waveform generation in the Fourier domain for LISA, comparing the performance of the new Python implementation with both the LALSuite-based implementation of \textsc{IMRPhenonmT} and \textsc{IMRPhenomX} families in LALSuite. For each model, we simulate 1000 random cases with mass ratios between [1-10] and total masses between $[10^6-10^7 M\odot]$ logarithmically distributed. Spin parameters are uniformly distributed in the $z$-component for the aligned spin models and uniformly distributed in polar coordinates for the precessing models. We set the frequency range to 0.1 mHz–0.1 Hz with a sampling time of 5 s. The total mass range is divided into 20 bins, and the average evaluation cost is computed within each bin.

Results indicate that the new Python implementation of \textsc{IMRPhenomT} provides significant computational advantages. Specifically, it outperforms the original C-based LAL implementation and achieves performance comparable to the Fourier-domain models on the CPU. In its GPU version, the Python implementation demonstrates a two-order-of-magnitude improvement over the LAL \textsc{IMRPhenomT} and a one-order-of-magnitude improvement over \textsc{IMRPhenomX}, the latter being the most computationally efficient waveform family available so far. It is also worth noting that our Python implementation of \textsc{IMRPhenomT} does not yet incorporate any nonuniform grid acceleration techniques, such as those used in \textsc{IMRPhenomX} with the Multibanding technique. Porting this technique to the time domain could reduce the computational cost even more.

The considerable speed-up and simplified structure provided by this new Python implementation motivates additional efforts to port the entire \textsc{IMRPhenomX} waveform family into Python as well. Ongoing work includes implementing new features such as the (2,0) mode with memory effects \cite{Maria} and eccentricity effects in the time domain \cite{Lluc} as well as in the Fourier domain \cite{Toni}. These enhancements aim to expand the versatility and precision of \textsc{IMRPhenom} models in gravitational wave analysis across a wider array of astrophysical scenarios, benefiting applications in both LISA and ground-based detectors.

\subsection{Time domain LISA response}

For current ground-based detectors, the response to incoming signals is computed simply multiplying the two polarizations by the well-known antenna patterns, which can be treated as static due to the short duration of the signals. However, in LISA, signals can span over a much longer time, necessitating a dynamical time-varying detector response. More importantly, the response from each link needs to be appropriately combined through TDI to suppress the laser frequency noise by several orders of magnitude. Without TDI, detecting gravitational wave signals under such significant noise would be nearly impossible. Working directly in the time domain allows for straightforward testing of various TDI configurations and the incorporation of realistic orbits with unequal-arm lengths, capabilities that are currently unattainable with the Fourier domain formalism.
Here we outline the method to compute LISA's time-dependent response, closely following the approach and notation established in the LISA Data Challenges (LDC) \cite{ldctools, radler, sangria}. 

LISA consists of a triangular configuration of three spacecraft, each interconnected by two laser beams in opposite directions. This arrangement yields three arms with six links, each link requiring an independent response computation. We denote these responses as $y_{slr}$, where $s$ identifies the sender spacecraft, $r$ the receiver and $l$ the link according to the convention in Fig.~1 of \cite{radler}.

To compute the LISA response, the positions of the sender and receiver ($\textbf{r}_{s}$, $\textbf{r}_{r}$) and the light travel time in each link ($L_{l}$) must be tracked over the whole observation period ($t_{\mathrm{obs}}$). These quantities can be obtained from multiple simulation codes of the LISA orbital dynamics and only need to be computed once. In Sec.~\ref{sec:unequal_arms} we will compare two types of orbits, one with equal-arms computed with the \texttt{AnalyticOrbit} class provided in the LISA Data Challenge tools \cite{ldctools} and a second one with realistic unequal-arm orbits provided by ESA \cite{esa_orbits}.

We define an orthonormal basis associated with the propagation vector \textbf{k}:
\begin{equation}
    \begin{aligned}
        \textbf{k} &= \big( -\cos\beta \cos\lambda, -\cos\beta \sin\lambda, -\sin\beta \big)\\
        \textbf{v} &= \big( -\sin\beta\cos\lambda, -\sin\beta\sin\lambda, \cos\beta \big)\\
        \textbf{u} &= \big( \sin\lambda, -\cos\lambda, 0 \big),
    \end{aligned}
\end{equation}
where \textbf{u}, \textbf{v} are known as the \textit{polarizations vectors} and $\lambda$, $\beta$ are the ecliptic longitude and latitude measured from the Solar System Barycenter (SSB). Then, for each link the response is computed as
\begin{equation}\label{eq:yslr}
    \begin{aligned}
        \textbf{n} &= \frac{\textbf{r}_{r} - \textbf{r}_{s}}{|\textbf{r}_{r} - \textbf{r}_{s}|}\\
        x_+ &= (\textbf{u}\cdot\textbf{n})^2 - (\textbf{v}\cdot\textbf{n})^2\\
        x_{\times} &= (\textbf{u}\cdot\textbf{n}) (\textbf{v}\cdot\textbf{n})\\
        t_s &= t_{\mathrm{obs}} - L_l - \textbf{k}\cdot{\textbf{r}_{s}}\\
        t_r &= t_{\mathrm{obs}} - \textbf{k}\cdot{\textbf{r}_{r}}\\
        y_{slr} &= \frac{x_+ \big(h_+(t_s) - h_{+}(t_r)\big) + x_{\times} \big(h_{\times}(t_s) - h_{\times}(t_r)\big)}{1 - \textbf{k}\cdot\textbf{n}}.
    \end{aligned}
\end{equation}

Next, we compute the TDI variables. The TDI technique is continually evolving, with multiple generations that utilize various data combinations and delayed times, each providing different levels of accuracy and noise cancellations. By operating in the time domain, one can implement any of these TDI generations. The LDC tools allow for the selection between the 1st (1.5) and the 2nd generations. The 1.5 generation Michelson X variable is defined as 
\begin{equation}
    \begin{aligned}
        X_{1.5} &= y_{1-32}(t - L_{3}- L_{2}- L_{-2}) \\
        &- y_{1-32}(t - L_{3})\\
        &+ y_{231}(t - L_{2}- L_{-2}) \\
        &- y_{231}(t) \\
        &+ y_{123}(t - L_{-2}) \\
        &- y_{123}(t - L_{-2} - L_{-3}- L_{3}) \\
        &+ y_{3-21}(t)\\
        &- y_{3-21}(t - L_{-3} - L_{3}),    
    \end{aligned}
\end{equation}
while the second generation reads as
\begin{equation}
    \begin{aligned}
        X_2 &= y_{1-32}(t - L_{3}- L_{2}- L_{-2})\\
        &- y_{1-32}(t - L_3) \\
        &+ y_{1-32}(t - L_{-2} - L_{2} - 2L_{3} - L_{-3}) \\
        &- y_{1-32}(t - 2L_3 - L_{-3} - 2L_{-2} - 2L_2)\\
        &+ y_{231}(t - L_2 - L_{-2}) \\
        &- y_{231}(t) \\
        &+ y_{231}(t - L_{-2} - L_{2} - L_{3} - L_{-3}) \\
        &- y_{231}(t - L_3 - L_{-3} - 2L_{-2} - 2L_2)\\
        &+ y_{123}(t - L_{-2}) \\
        &- y_{123}(t - L_{-2} - L_{-3} - L_3) \\
        &+ y_{123}(t - 2L_{-2} - L_{2} - L_{3} - L_{-3}) \\
        &- y_{123}(t - L_3 - L_{-3} - 2L_{-2} - L_2)\\
        &+ y_{3-21}(t) \\
        &- y_{3-21}(t - L_{-3} - L_3) \\
        &+ y_{3-21}(t - L_{-2} - L_{2} - 2L_{3} - 2L_{-3}) \\
        &- y_{3-21}(t - L_3 - L_{-3} - L_{-2} - L_2).
    \end{aligned}
\end{equation}
The other two Michelson variables $Y$, $Z$ are obtained by simple and double cyclic permutations of the $slr$ indices respectively. Independent channels can be constructed under the premise of identical and uncorrelated noise across the detector arms as
\begin{equation}
    A=\frac{Z-X}{\sqrt{2}}, \:E=\frac{X-2Y+Z}{\sqrt{6}}, \:T=\frac{X+Y+Z}{\sqrt{3}}.
\end{equation}
The TDI calculations necessitate the evaluation of the arm responses at delayed times. For our GPU version, we will use linear interpolation due to the lack of a fast spline implementation in the \texttt{cupy} library. Given that the $y_{slr}$ responses are finely sampled in time, linear interpolation is adequate enough.

Once the three TDI channels (AET) have been obtained, we perform a Fourier transform on them and compute the log-likelihood as \footnote{We omit here the constant term arising from the product of the data $(d|d)$. The complete definition of the log-likelihood is $\log \mathcal{L}= - \frac{1}{2} ( h - d | h - d)$.}
\begin{align}
\begin{split}
    \log \mathcal{L} = - \big(
    &(\tilde{A}_{inj}|\tilde{A}) - \frac{1}{2}(\tilde{A}|\tilde{A}) \\
    + &(\tilde{E}_{inj}|\tilde{E}) - \frac{1}{2}(\tilde{E}|\tilde{E}) \\+ &(\tilde{T}_{inj}|\tilde{T}) - \frac{1}{2}(\tilde{T}|\tilde{T}) 
    \big).
\end{split}
\end{align}
The subscript \textit{inj} designates the injected signal, which plays the role of real data, while the variables without subscripts refer to the templates. The inner product $(a|b)$ denotes the standard matched-filter inner product
\begin{equation}
    (\tilde{a}|\tilde{b}) = 4 \mathrm{Re} \int_0^\infty \frac{\tilde{a}(f) \tilde{b}^*(f)}{S_n(f)}df.
\end{equation}
In practice, the integral is evaluated over the frequency range of the detector. $S_n$ is the power spectral density (PSD), and we adopt the analytical noise model employed in the Sangria Data Challenge \cite{sangria}. The noise contributions from the acceleration and optical metrology systems are
\begin{equation}
    \begin{aligned}
        S_{\mathrm{acc}} &= \left(2.4\cdot 10^{-12}\right)^2 \:\left(1 + \left(\frac{4\cdot10^{-4}}{f}\right)^2\right) \left(1 + \left(\frac{f}{0.008}\right)^4\right) \times \\&\;\;\; \left(2 \pi f\right)^{-4} \left(\frac{2 \pi f}{c}\right)^2\\ 
        S_{\mathrm{op}} &= \left(7.9\: 10^{-12}\right)^2 \left(1 + \left(\frac{0.002}{f}\right)^4\right) \left(\frac{2 \pi f}{c}\right)^2.
    \end{aligned}
\end{equation}
Then the PSDs for each TDI channel become
\begin{equation}
    \begin{aligned}
        S_{n,AE} &= 8 \sin^2(2\pi f) (2 \: S_{\mathrm{acc}}  (3 + 2 \cos(2\pi f) + \cos(4\pi f)) \\ &+ S_{\mathrm{op}} (2 + \cos(2\pi f)))\\
        S_{n,T} &= 16\: S_{\mathrm{op}} \left(1 - \cos(2\pi f)\right) \sin^2(2\pi f) \\&+ 128 \:S_{\mathrm{acc}} \sin^2(2\pi f) \sin^4(\pi f).
    \end{aligned}
\end{equation}

\subsection{Injections setup}
In the most general scenario, one would use the full LISA frequency range $10^{-4}-0.1$ Hz, which translates to a sampling time interval of $\Delta t = 1/(2f_{max}) = 5s$. Here we will generate the waveforms with a minimum frequency of $10^{-4}$ Hz, but sample instead every $\Delta t = 10s$ and then linearly interpolate them to generate the arm and TDI responses with a sampling time of $5$ s for a faster evaluation. Once the Fourier transform is performed, we will restrict the frequency range to $3\cdot 10^{-4} - 0.02$ Hz to expedite the inner products calculation. The injected signal will represent an MBHB with a duration of approximately two months and will be merged well below the maximum frequency of 0.02 Hz. We will inject this system with and without precessing spins under zero-noise\footnote{The time series representing the injected signal, designed to mimic the detector's data, comprises solely the gravitational wave template, with no noise time series added.} conditions in the signal.
The parameters to be injected, along with the corresponding uniform prior ranges, are detailed in Tab.~\ref{tab:injection}.

For the injection and recovery processes, we will utilize the same waveform model and same TDI configurations. Bayesian inference will be executed using the \textsc{Bilby} package \cite{bilby_paper}, which is extensively used for analyzing data in the LVK collaboration. Notably, no modifications to the \textsc{Bilby} package are necessary to interface with our new waveform generation and LISA response calculations, as this functionality has been integrated into the waveform generator. A preliminary profiling test for the aligned-spin case revealed that the Fourier transform, arm response, and TDI calculation contribute approximately to the 10\% of the total computational cost. In contrast, waveform model generation accounts for 20-40\%, while likelihood calculation represents 50-70\%. This substantial computational cost associated with the likelihood calculation has been identified, and a solution will be provided in a future update of our code. The goal is to ensure that the primary computational cost is attributed solely to the waveform model generation.

We conducted preliminary test runs in which a reduced number of parameters were sampled, and these runs successfully recovered all the parameters with Gaussian-shaped posteriors. However, as we increased the number of parameters in the precessing case,
the three main samplers we tested (\textsc{ptemcee} \cite{ptemcee}, \textsc{dynesty} \cite{dynesty}, and \textsc{multinest} \cite{multinest}) did not converge within a time frame of a few days, due to the high computational cost and the high dimensionality of the parameter space.
In order to speed up the results, the injections presented in the next section will not commence with a random sampling across the parameter space but will start from a region near the injected values based on Fisher estimates. This \textit{Fisher initialization} method generates random starting points drawn from a multivariate Gaussian distribution where the mean is given by the injected values and the covariance matrix by the inverse of the Fisher matrix. This initialization method is also used in other pipelines like \texttt{lisabeta} \cite{Marsat_2021}.
The Fisher matrix is computed from the derivatives of the likelihood function, using first-order finite differences as implemented in the \texttt{calculate\_FIM} function within \textsc{Bilby}.

The number of initial points provided to the sampler corresponds to the number of live points for \textsc{dynesty} and the product of the number of walkers and temperatures for \textsc{ptemcee}. Despite this initialization, we found that the \textsc{dynesty} sampler struggled more than \textsc{ptemcee} to achieve convergence, prompting us to focus exclusively on the results obtained with \textsc{ptemcee}. For our analysis, we utilized 28 walkers and 4 temperatures for all the runs, executing the computations on a single Nvidia Tesla V100 GPU.

Another approach to enhance the convergence speed of the sampler is to employ \textit{Fisher priors}, which utilize the diagonal values of the covariance matrix as standard deviations for univariate Gaussian distributions assigned as priors for each parameter. Both methods necessitate prior knowledge of the true parameters or at least values in close proximity to them. This scenario is analogous to analyses utilizing heterodyne likelihood \cite{Cornish_2021} or relative binning \cite{Zackay_2018, Leslie_2021, Krishna_2023} algorithms, both of which require a fiducial point situated near the maximum likelihood estimate. Such a point can often be derived from the best-fitting template obtained via a matched-filter search, where the injected signal is compared against a pre-generated bank of gravitational wave templates. In-depth investigations are required to assess how effectively these estimates perform across the parameter space, the necessary proximity to the true values, and whether estimates derived from dominant-mode, non-precessing template banks can be applied to analyses involving subdominant harmonics and precession.

In the following section, we demonstrate that using Fisher initialization allows us to accurately recover all injected parameters within a few hours. To validate our estimates, we will compare our Bayesian posteriors with those derived from a full Fisher analysis. The Fisher analysis serves as a useful approximation for parameter estimation, achieving exact results in the limit of infinite SNR \cite{Cutler_Flanagan, Cutler_Vallisneri, Vallisneri_2008}. Consequently, as the SNR increases, we would expect our Bayesian inference posteriors to approach the Fisher results. At finite SNR, our posteriors should be less informative, appearing ``wider" compared to those from the Fisher analysis. This relationship will be examined in Sec.~\ref{sec:fisher}. The Fisher posteriors are obtained by drawing $10^4$ samples from the Gaussian multivariate distribution previously described. In future work, we will investigate the potential for running without sampler initialization by further optimizing the likelihood evaluation process. For now, the configuration presented here serves as a critical testbed, providing valuable insights into LISA's capabilities for precise parameter measurements across various scenarios.

\begin{table}[ht]
\setlength{\tabcolsep}{ 6pt}  
\renewcommand{\arraystretch}{1.2}  
\centering
\begin{tabular}{ l r c }
Parameter & Injected & Prior range \\ \hline 
mass1, $m_1\: [10^6 M_\odot]$ & 1.0 & (0.99, 1.01)\\ 
mass2, $m_2\: [10^6 M_\odot]$ & 0.7 & (0.69, 0.71)\\ 
spin1\_norm, $a_1$ & 0.7 & (0.0, 1.0)\\ 
spin1\_tilt, $\theta_1$ [deg] & 40.1 & (0.0, 180.0)\\ 
spin1\_phi, $\phi_1$ [deg] & 40.1 & (-180.0, 180.0)\\ 
spin2\_norm, $a_2$ & 0.4 & (0.0, 1.0)\\ 
spin2\_tilt, $\theta_2$ [deg] & 22.9 & (0.0, 180.0)\\ 
spin2\_phi, $\phi_2$ [deg] & 22.9 & (-180.0, 180.0)\\ 
spin1z (aligned-spin) & 0.7 & (-1.0, 1.0)\\ 
spin2z (aligned-spin) & 0.4 & (-1.0, 1.0)\\ 
distance, D [Gpc] & 20.0 & (19.0, 21.0)\\ 
inclination, $\iota$ [deg] & 149.0 & (0.0, 180.0)\\ 
phiRef, $\phi$ [deg] & 229.2 & (0.0, 360.0)\\ 
EclipticLatitude, $\beta$ [deg] & -34.4 & (-180.0, 180.0)\\ 
EclipticLongitude, $\lambda$ [deg] & 34.4 & (0.0, 360.0)\\ 
polarization\_angle, $\psi$ [deg] & 22.9 & (-180.0, 180.0)\\ 
minimum frequency [Hz] & $10^{-4}$ & -\\ 
reference frequency [Hz] & $10^{-4}$ & -\\ 
sampling time $\Delta t$ [s] & 10.0 & -\\ 
duration [days] & 61 & -\\
SNR & $\sim$2800 & -\\
\end{tabular}
\caption{Injected values and range of uniform priors both for the aligned spin and precessing systems. The SNR takes the approximate average value between four injected signals (aligned spin and precessing with and without subdominant harmonics).}
\label{tab:injection}
\end{table}

\section{Results} \label{sec:results}
We present a set of zero-noise injections to assess the LISA’s capacity to measure source parameters for both aligned-spin and precessing systems. We examine the effect of including subdominant harmonics on parameter recovery, compare the Bayesian posterior distributions against the approximate Fisher analyses, and explore the influence of unequal-arms, different TDI channels and TDI generations on the results. These comparisons are among the first to utilize full Bayesian inference, providing valuable insights into the benefits of incorporating additional physical effects in waveform models and understanding how various LISA configurations may impact the accuracy of parameter recovery.

Bayesian inference is carried out employing the \textsc{Bilby} package in serial mode, with the \textsc{ptemcee} sampler initialized via Fisher matrices to expedite convergence. All the runs employ an equal-arm configuration except those in Sec.~\ref{sec:unequal_arms}.


In this initial study, both the signal injection and parameter recovery are performed with exactly the same waveform model and LISA configuration. A more realistic approach would use a fixed injected signal (such as an NR simulation) and then conduct parameter recovery with various waveform models (e.g., \textsc{IMRPhenomT}, \textsc{IMRPhenomTHM}, \textsc{IMRPhenomTPHM}...) and different LISA settings. However, due to the higher computational demand of this setup, such analysis is deferred to future work.


In the following subsections, we present multivariate posterior distributions using standard corner plots. The 2D distributions display the 1-$\sigma$, 2-$\sigma$, and 3-$\sigma$ equivalent contours, corresponding to 39\%, 86\%, and 99\% of the enclosed probability volume, respectively. In the 1D marginal distributions along the diagonal of the plots, dashed vertical lines indicate the $1-\sigma$ credible intervals (quantiles of 0.16 and 0.84), while solid black lines denote the injected parameter values.

The median values, the 90\% credible intervals, and the relative errors with respect to the injected values\footnote{Relative errors are computed dividing the width of the 90\% credible intervals by the injected value.} are summarized in Tabs.~\ref{tab:22_vs_HM} and \ref{tab:tdis}, alongside the total number of samples and sampling times for each run.

\subsection{Effect of subdominant harmonics}\label{sec:subeffect}

\begin{figure*}[ht]
  \centering
  \includegraphics[width=\textwidth]{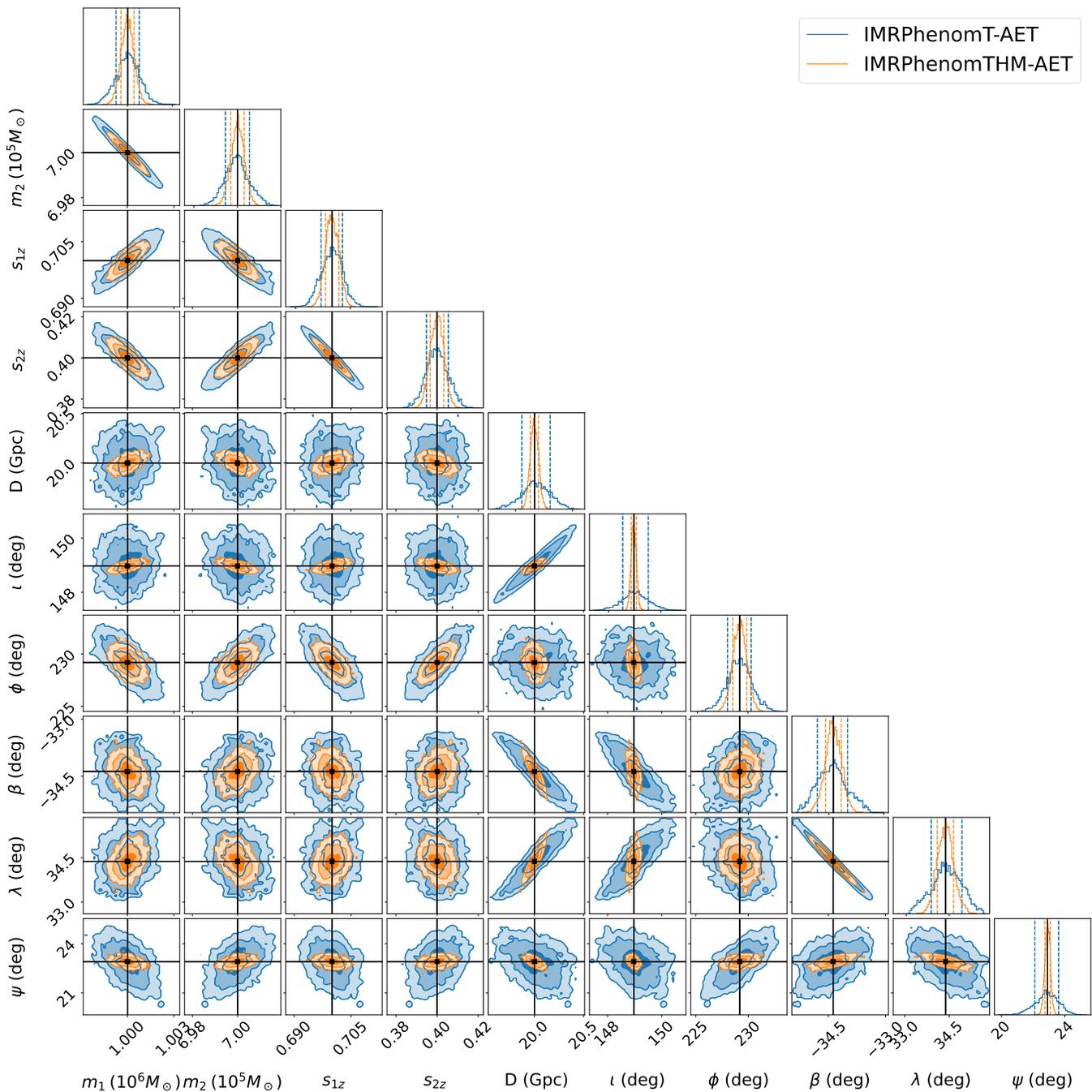} 
  \caption{Comparison of the effect of including subdominant harmonics for the aligned-spin injection. In blue the dominant mode model and in orange the subdominant modes one. Both employ the three TDI channels (AET).}
  \label{fig:22_vs_HM_AS}
\end{figure*}

\begin{figure*}[ht]
  \centering
  \includegraphics[width=\textwidth]{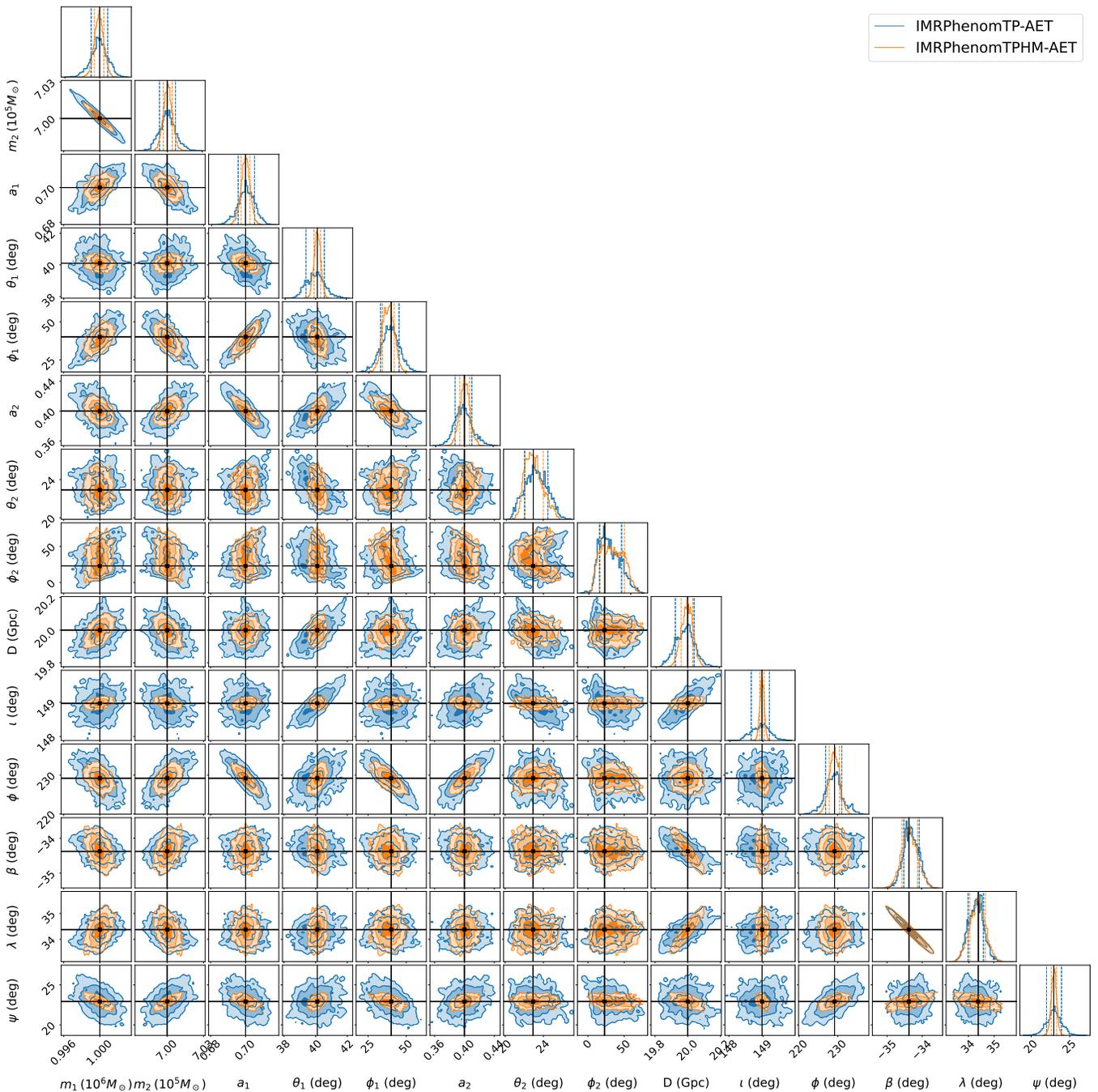} 
  \caption{Comparison of the effect of including subdominant harmonics in the co-precessing frame for the precessing injection. In blue the dominant mode model and in orange the subdominant modes one. Both employ the three TDI channels (AET).}
  \label{fig:22_vs_HM_prec}
\end{figure*}

Incorporating additional physical effects in gravitational waveforms generally enhances the accuracy of source parameter estimation by breaking degeneracies between parameters. In particular, subdominant harmonics are crucial to break the degeneracy between inclination and distance \cite{phenomhm} and between the polarization angle and the reference phase. 

Distance impacts the waveform by overally scaling its amplitude. For the dominant mode, the inclination angle entering in the spin-2 weighted spherical harmonics ($Y_{22}$, $Y_{2-2}$) is also an overall scaling factor, therefore degenerated with the distance. With subdominant harmonics included, the inclination affects each individual harmonic, changing their relative contribution to the multimode waveform, thus breaking the degeneracy.
Similarly, the polarization angle $\psi$ is an overall phase offset (multiplying the waveform $h_+ -i h_\times$ by $e^{i 2 \psi}$). For a waveform containing only the dominant mode, this resembles the influence of the reference phase, as it enters through the complex exponential of the $Y_{lm}$ as $e^{i m \phi}$ and $m=2$. With subdominant harmonics, the reference phase affects each harmonic differently, thus breaking the degeneracy similarly to the previous case.

These two degeneracy breakings can be observed very prominently in Fig.~\ref{fig:22_vs_HM_AS} for the aligned-spin case. The measurements of distance, inclination, reference phase and polarization angle are much more stringent with subdominant harmonics. This observation aligns with the findings reported in \cite{Marsat_2021} also in the context of LISA parameter estimation.

The accuracy of the rest of the parameters also improves including subdominant modes. In Tab.~\ref{tab:22_vs_HM} we show all the recovered parameters with their 90\% confidence intervals and relative errors with respect to the injected values. Notably, unlike current ground-based detectors, LISA demonstrates the ability to accurately measure individual spin components, which is achieved at the 1\% level. This capability is crucial for population analysis and aids in discerning between different formation channels. Furthermore, we emphasize the importance of incorporating subdominant harmonics to achieve a more precise determination of the source's sky location, which is essential for multimessenger follow-up observations.

In Fig.~\ref{fig:22_vs_HM_prec} we present the equivalent results for the precessing case. This comprises the first comparison of how waveform models with subdominant harmonics impact the recovery of parameters in LISA observations for precessing systems. It is important to remember that for precessing models, the inclusion of dominant and subdominant harmonics refers to those in the co-precessing frame, and that after performing the rotation to the inertial frame, the model approximates all the harmonics with the same $l$. For example, \textsc{IMRPhenomTP}, incorporates only the 22 mode in the co-precessing frame, but in the final inertial frame approximates all the $l=2$ modes. This means that the dominant mode model \textsc{IMRPhenomTP} has also some capacity to break degeneracies and might explain why the inclusion of subdominant harmonics here has a milder effect compared to the aligned-spin case. 

While the inclusion of subdominant modes leads to more restrictive priors for most parameters, the posteriors for sky location remain nearly unchanged. The recovery of individual spin components is significantly improved compared to ground-based detectors, with the two norms and the tilt angle of the primary spin being estimated at the 1\% level. However, the azimuthal angles exhibit poorer recovery, likely due to the use of the NNLO single-spin prescription for the Euler angles that characterize precession. In this prescription, the azimuthal angles do not substantially affect the waveform's shape, complicating the sampler's ability to find the correct solution. For future investigations, it will be essential to explore more sophisticated modelling approaches that accurately account for double-spin contributions, thereby enhancing the overall parameter recovery process in LISA observations.

When comparing results with and without precession, we observe no substantial improvement in the accuracy of recovered parameters (see Tab.~\ref{tab:22_vs_HM}). Initially, we hypothesized that this limited impact might stem from the injected system's orientation, as it was close to face-off ($\theta_{JN}=153^\circ$); thus, we conducted additional analyses on an edge-on system but again observed no significant improvement.
In fact, if we compare the posteriors from the Fisher analysis in the next section, we see that parameters like masses, distance and sky location have a noticeable accuracy improvement when including precession. This could suggest a need for enhanced sampling configurations to adequately navigate the higher-dimensional parameter space introduced by precessing systems. In this study, the sampler settings were held the same across all the runs, without in-depth tuning for precessing cases, which may limit parameter space exploration. Furthermore, as we are comparing recoveries of different injected signals (a non-precessing one versus a precessing one) our initial expectation that precession would consistently enhance parameter recovery may not fully hold in this context. A comprehensive, systematic investigation is planned for future work. Here, our main aim is to showcase illustrative examples of potential outcomes with the newly developed framework.

\subsection{Comparison against Fisher analysis}\label{sec:fisher}

\begin{figure*}[ht]
  \centering
  \includegraphics[width=\textwidth]{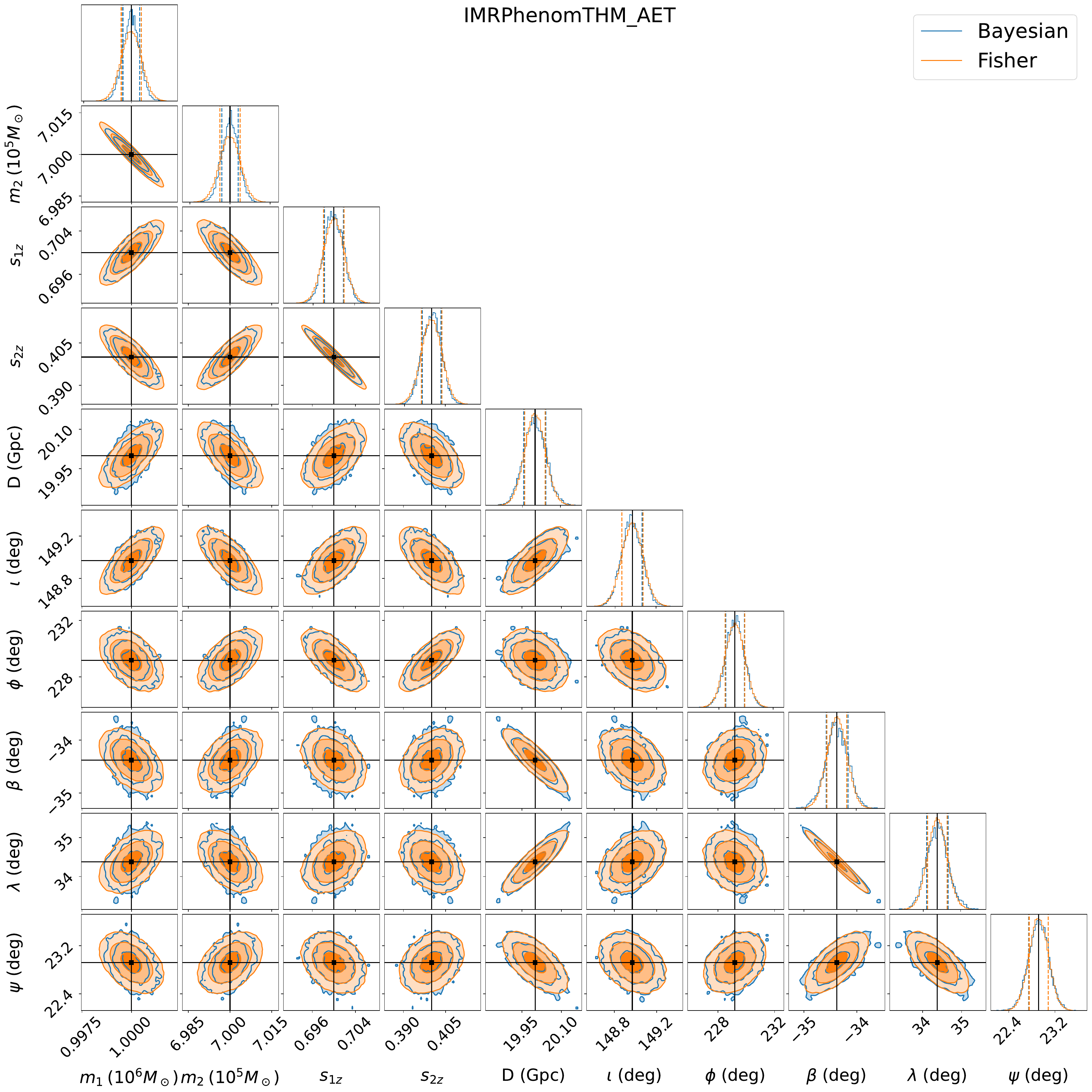} 
  \caption{Comparison of Bayesian inference and Fisher estimators for an aligned spins injection with subdominant harmonics.}
  \label{fig:Fisher_AS}
\end{figure*}

\begin{figure*}[ht]
  \centering
  \includegraphics[width=\textwidth]{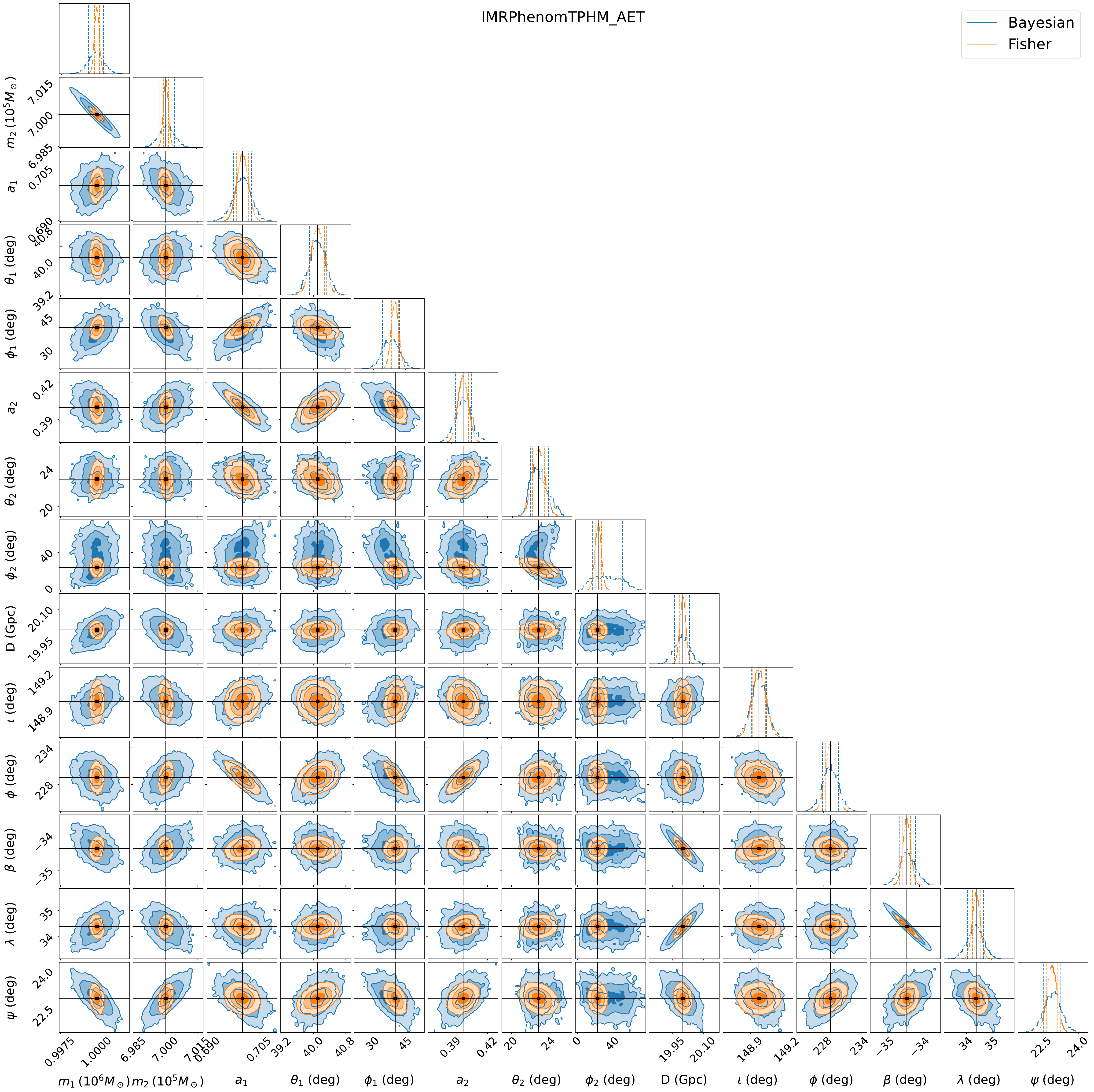} 
  \caption{Comparison of Bayesian inference and Fisher estimators for a precessing injection with subdominant harmonics.}
  \label{fig:Fisher_prec}
\end{figure*}

In order to assess the consistency of the results presented in the previous section, we compare the Bayesian posteriors to those obtained from a complete Fisher analysis for cases including subdominant harmonics. Given the high SNR of the system, the Bayesian posteriors are expected to approximate the Fisher estimates. 

In Fig.~\ref{fig:Fisher_AS}, we show results for the aligned spin scenario, observing good agreement between the two methods. For the mass parameters, however, the Bayesian posteriors appear slightly narrower than those from the Fisher analysis, potentially due to an artifact of the Fisher initialization. This effect might arise if the sampler locates the peak too quickly, resulting in incomplete sampling of the posterior ellipse. In our other analyses, such as those focused solely on the 22 mode or employing only the AE channels, the Bayesian posteriors appear wider than those obtained from the Fisher analysis, as expected. Additionally, it is important to note that the Fisher matrix formalism is particularly sensitive to numerical inaccuracies, which can lead to scenarios where the inverse covariance matrix is not positive semidefinite. Although we did not encounter this issue in the examples presented here, we should be aware that the Fisher posteriors could be affected by numerical errors.

In the precessing case illustrated in Fig.~\ref{fig:Fisher_prec}, we observe that the posteriors exhibit notable differences, with the Bayesian posteriors consistently wider than those derived from the Fisher analysis, which is anticipated. This widening may be attributed to the increased dimensionality associated with precessing systems, requiring the sampler to explore a larger region of the parameter space before locating the correct peak. We should remind that, despite employing Fisher initialization, the priors are wide and allow the sampler to explore regions outside the Fisher ellipse, potentially influencing the width of the resulting posteriors.

\subsection{Comparison equal vs unequal-arms}\label{sec:unequal_arms}
We investigate the impact of incorporating realistic orbits with unequal-arms for the LISA constellation, as provided by ESA \cite{esa_orbits}, compared to scenarios assuming equal-arms. The equal-arm cases correspond to those discussed in Sec.~\ref{sec:subeffect}. For the unequal-arm configuration, we adopt new orbits while keeping the PSDs unchanged. The use of more realistic PSDs will be postponed for future work.

\begin{figure*}[ht]
  \centering
  \includegraphics[width=\textwidth]{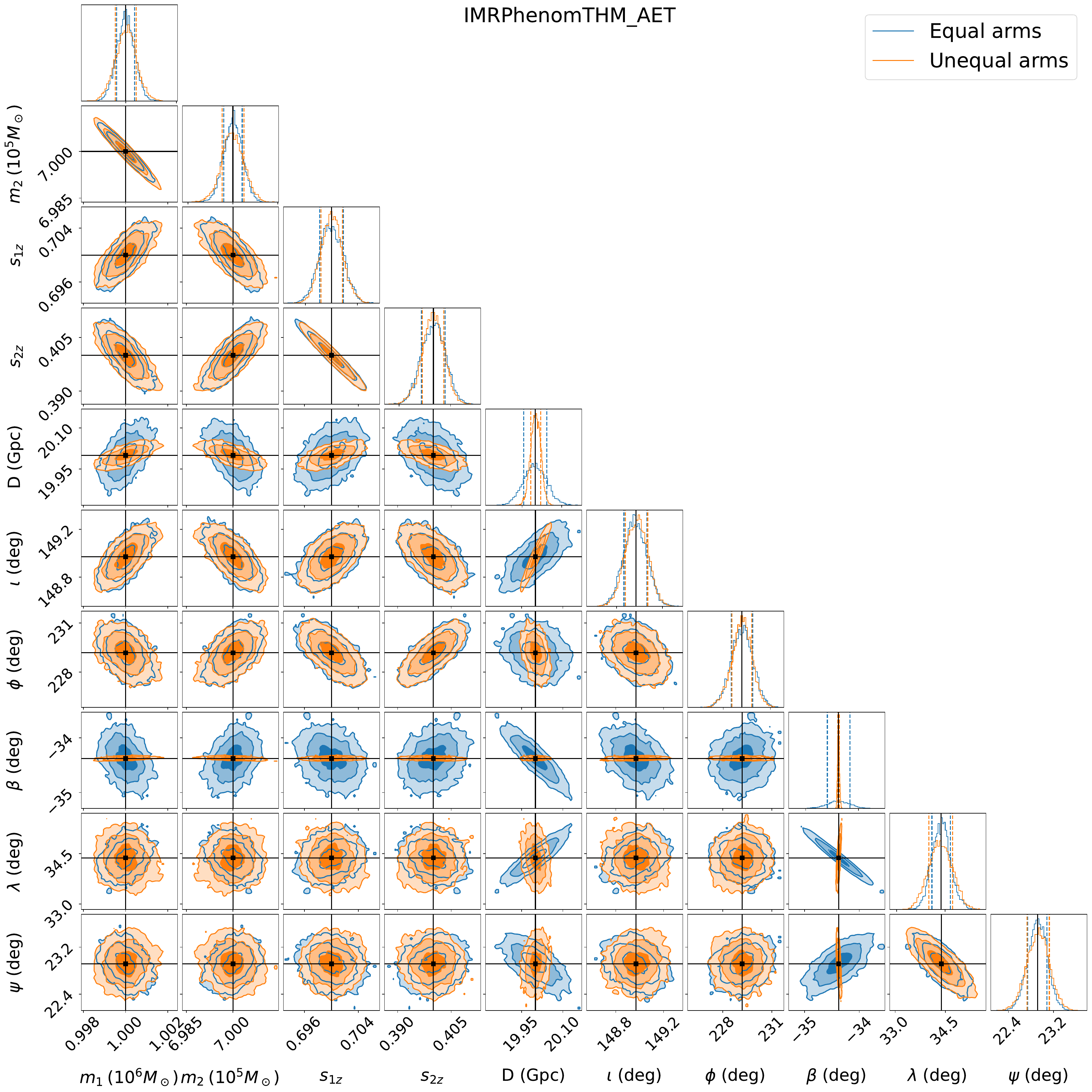} 
  \caption{Posterior distribution comparison between equal and unequal-arm orbits for \textsc{IMRPhenomTHM} with 3 TDI channels.}
  \label{fig:THM_unequal}
\end{figure*}

\begin{figure*}[ht]
  \centering
  \includegraphics[width=\textwidth]{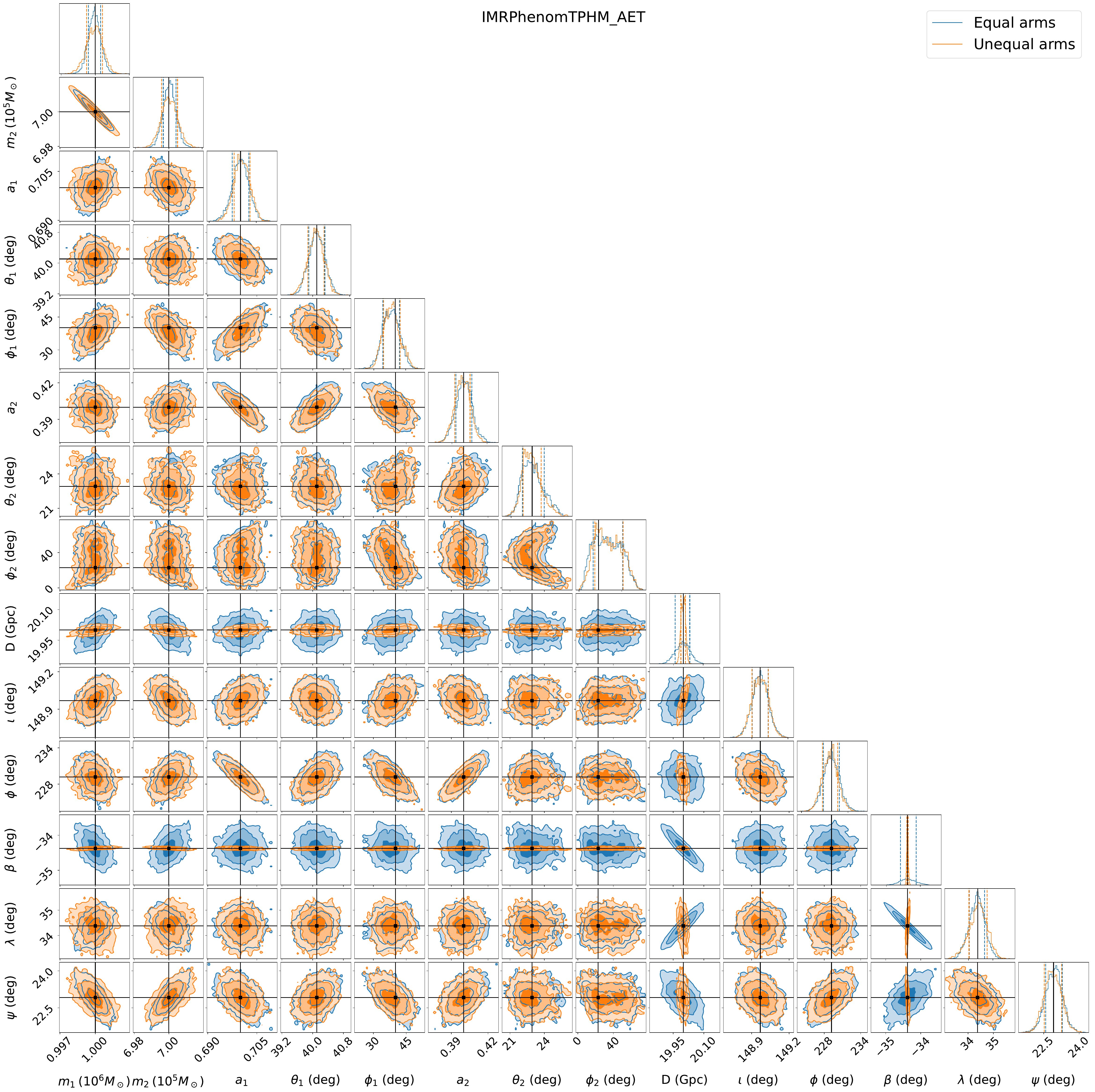} 
  \caption{Posterior distribution comparison between equal and unequal-arm orbits for \textsc{IMRPhenomTPHM} with 3 TDI channels.}
  \label{fig:TPHM_unequal}
\end{figure*}

Figs.~\ref{fig:THM_unequal}, \ref{fig:TPHM_unequal} show the results for the aligned-spin and precessing cases respectively, both including subdominant harmonics. It can be seen that the unequal-arm configuration leads to improved recovery of the ecliptic latitude and distance.

Variations in the LISA arm lengths are intrinsically linked to the sky location of the source (see Eq.~\ref{eq:yslr}), meaning that additional amplitude and phase time-dependent modulations introduced by LISA’s breathing motion help to resolve degeneracies and improve measurement precision. The improved sky location together with the amplitude modulations helps to better constrain the distance. The angular resolution of LISA depends on sky location \cite{Cutler_1998}, and changes in ecliptic latitude and longitude influence the signal differently depending on the specific sky location parameters. For the values considered here, a shift in latitude of $1^\circ$ causes amplitude and phase variations in the TDI channels an order of magnitude larger than a shift in longitude, highlighting greater sensitivity in the latitude. This role may vary across different sky locations, with the longitude becoming more sensitive. A comprehensive study of this behaviour across the celestial sphere is deferred to future work. Additionally, a more detailed analysis of how variations in orbital trajectories influence the recovery of the parameters is also left for subsequent investigation.

The work of \cite{Katz_2022} also compared the effect of unequal-arms with Bayesian inference but on galactic binaries. This study is the first to examine the effect on MBHB with generic spins, representing a significant advancement toward more realistic analyses with LISA. Future work will focus on more systematic investigations of unequal-arms, as this area remains largely unexplored. 

\subsection{Comparison of TDI channels}
\begin{figure*}[ht]
  \centering
  \includegraphics[width=\textwidth]{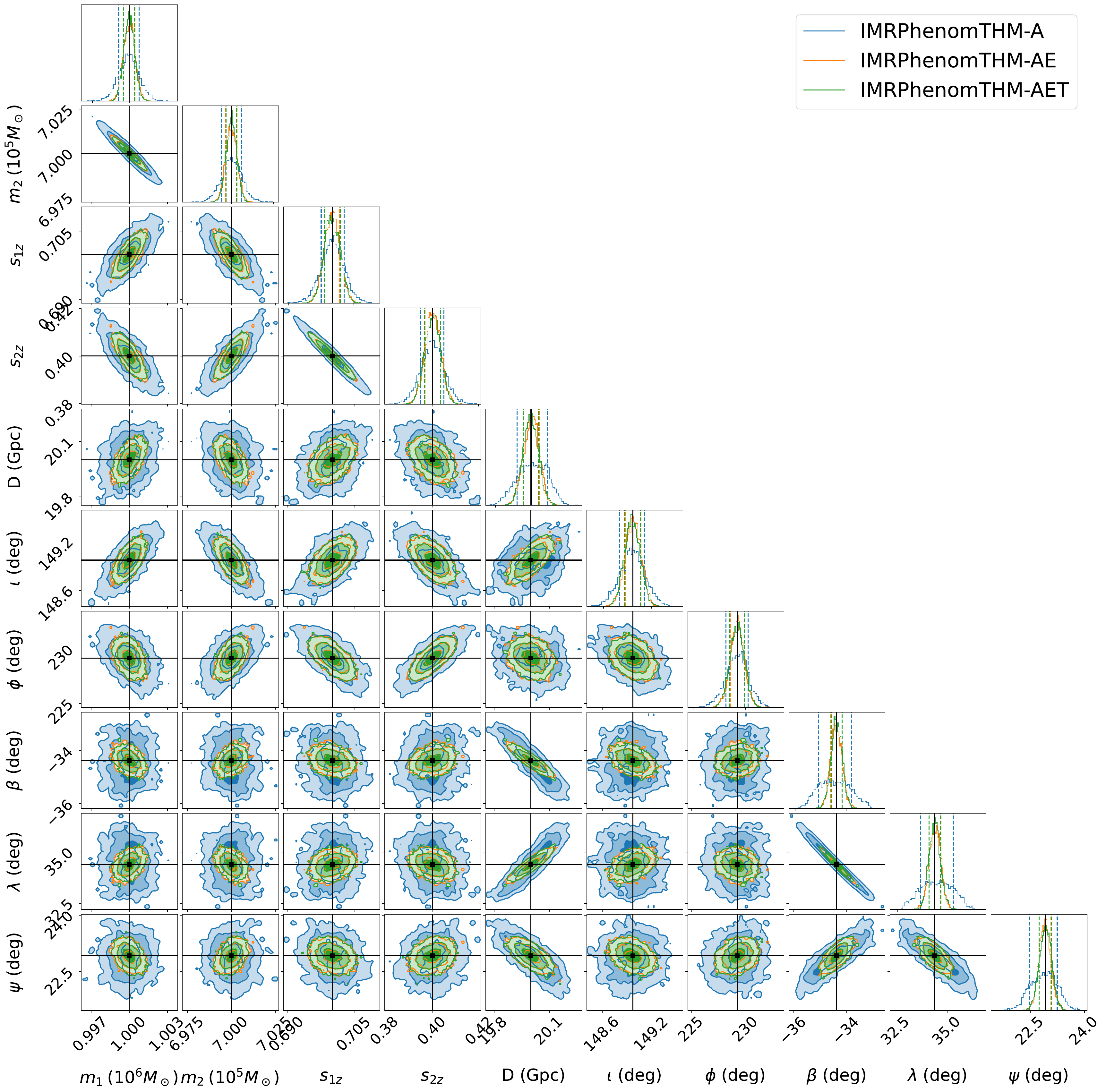} 
  \caption{Comparison of the effect of considering different TDI channels on the recovered parameters in an aligned-spin system. The green injection corresponds to the orange one in Fig.~\ref{fig:22_vs_HM_AS}.}
  \label{fig:TDI_channels_AS}
\end{figure*}

\begin{figure*}[ht]
  \centering
  \includegraphics[width=\textwidth]{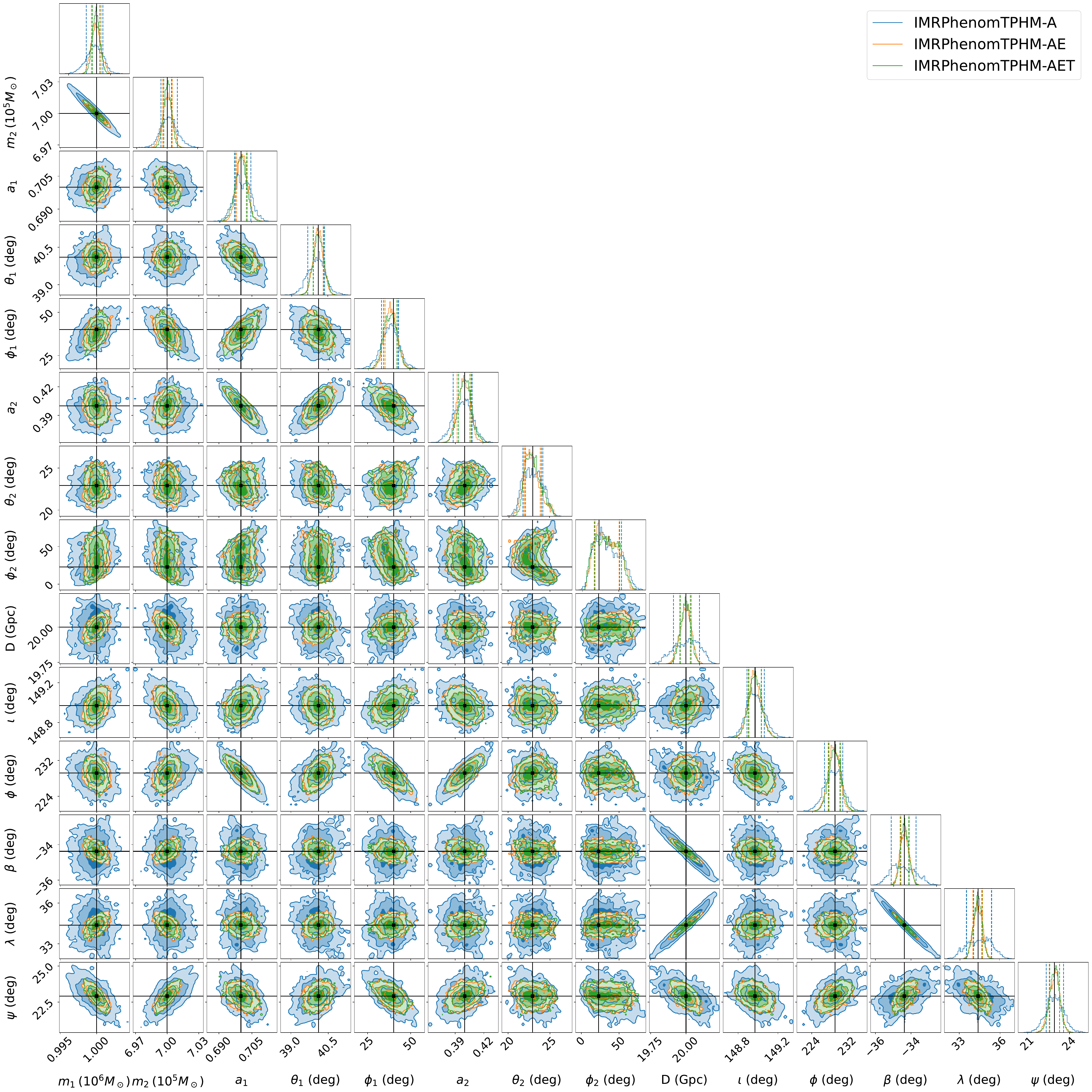} 
  \caption{Comparison of the effect of considering different TDI channels on the recovered parameters in a precessing system. The green injection corresponds to the orange one in Fig.~\ref{fig:22_vs_HM_prec}.}
  \label{fig:TDI_channels_prec}
\end{figure*}

In this section, we present the first analysis of how different TDI channel combinations affect full Bayesian inference in LISA observations. Figs.~\ref{fig:TDI_channels_AS} and \ref{fig:TDI_channels_prec} display the results for aligned spin and precessing cases, respectively. For each case, injections and recoveries were performed using one (A), two (AE) and three (AET) TDI channels. Channels A and E are constructed to capture the majority of the gravitational wave signal, while the T channel predominantly captures instrumental noise. Excluding the E channel results in broader, less confident posteriors, as it disregards substantial signal information. In contrast, excluding the T channel has little effect under zero-noise injections, though this channel is likely to play a larger role in analyses that include realistic noise, a topic that will be explored in future studies.

\subsection{Comparison of TDI generations}
We present a novel comparison of parameter estimation posteriors between the first (1.5) and second TDI generations. To our knowledge, only \cite{Wang_2024} has examined this impact but on non-spinning signals. Our work opens the opportunity to explicitly analyze the effect of TDI generations on inference accuracy and assess the potential value of advancing TDI development. We perform the injections and recoveries using three TDI channels (AET) computed for both the first (1.5) and second TDI generations, in aligned spin and precessing cases. We do not display the plots for this case since there are no significant differences between the two TDI generations, but the recovered values are provided in Tab.~\ref{tab:tdis}. For the aligned-spin injection, the posterior distributions overlap almost completely, while for the precessing case, there is a minor difference in the azimuthal angle of the secondary spin. This discrepancy likely reflects the inherent challenges for the sampler in accurately measuring this parameter, as discussed in Sec.~\ref{sec:subeffect}, rather than an effect of the TDI generations themselves. Notably, in the precessing case, although the posterior distributions remain similar, the sampler required a significantly greater computational time ($\sim$ 9h) and collected fewer samples overall (see Tables~\ref{tab:22_vs_HM} and \ref{tab:tdis}). The increased computational demand associated with the second TDI generation might be due to the more complex structure of its signal response.

Based on these results, we can conclude that, for zero-noise injections, the differences between TDI generations are not significant. This is not surprising, as each TDI generation uses different methods to suppress laser noise in the signal, which has little impact in the absence of injected noise. This result will probably change in the presence of noise, which we will explore in future work.




\begin{table*}[ht]
\setlength{\tabcolsep}{ 2pt}  
\renewcommand{\arraystretch}{1.4}  
\hspace*{-1cm}
\begin{tabular}{ l | r r r r r r }
\hline
Parameter & T-AET & THM-AET & UA-THM-AET & TP-AET & TPHM-AET & UA-TPHM-AET \\
\hline
$m_1\: (10^6M_\odot)$ & ${1.0000}_{-0.0014}^{+0.0012}$ (0.26\%) & ${1.0000}_{-0.0007}^{+0.0007}$ (0.15\%) & ${1.0000}_{-0.0009}^{+0.0009}$ (0.17\%) & ${1.0000}_{-0.0017}^{+0.0016}$ (0.33\%) & ${0.9999}_{-0.0009}^{+0.0009}$ (0.18\%) & ${1.0000}_{-0.0012}^{+0.0011}$ (0.24\%) \\ 
$m_2\: (10^5M_\odot)$ & ${6.9998}_{-0.0088}^{+0.0097}$ (0.26\%) & ${7.0001}_{-0.0052}^{+0.0051}$ (0.15\%) & ${6.9999}_{-0.0061}^{+0.0062}$ (0.18\%) & ${7.0002}_{-0.0113}^{+0.0123}$ (0.34\%) & ${7.0005}_{-0.0064}^{+0.0064}$ (0.18\%) & ${7.0003}_{-0.0078}^{+0.0084}$ (0.23\%) \\ 
$a_1$ &  -  &  -  &  -  & ${0.701}_{-0.009}^{+0.008}$ (2.4\%) & ${0.700}_{-0.005}^{+0.004}$ (1.3\%) & ${0.700}_{-0.004}^{+0.004}$ (1.2\%) \\ 
$\theta_1$ (deg) &  -  &  -  &  -  & ${40.0}_{-0.9}^{+1.0}$ (4.8\%) & ${40.1}_{-0.4}^{+0.4}$ (1.8\%) & ${40.1}_{-0.4}^{+0.3}$ (1.8\%) \\ 
$\phi_1$ (deg) &  -  &  -  &  -  & ${39.3}_{-10.5}^{+10.3}$ (51.9\%) & ${38.2}_{-6.3}^{+6.3}$ (31.5\%) & ${38.3}_{-6.1}^{+7.0}$ (32.6\%) \\ 
$a_2$ &  -  &  -  &  -  & ${0.398}_{-0.018}^{+0.023}$ (10.1\%) & ${0.400}_{-0.012}^{+0.012}$ (5.8\%) & ${0.399}_{-0.011}^{+0.010}$ (5.3\%) \\ 
$\theta_2$ (deg) &  -  &  -  &  -  & ${23.2}_{-2.0}^{+2.0}$ (17.3\%) & ${22.9}_{-1.3}^{+1.9}$ (13.6\%) & ${22.8}_{-1.1}^{+1.7}$ (12.4\%) \\ 
$\phi_2$ (deg) &  -  &  -  &  -  & ${28.3}_{-18.5}^{+28.6}$ (205.6\%) & ${33.0}_{-22.7}^{+25.6}$ (211.1\%) & ${33.3}_{-21.2}^{+26.0}$ (205.8\%) \\ 
$s_{1z}$ & ${0.700}_{-0.005}^{+0.004}$ (1.3\%) & ${0.700}_{-0.003}^{+0.003}$ (0.9\%) & ${0.700}_{-0.003}^{+0.003}$ (0.8\%) &  -  &  -  &  -  \\ 
$s_{2z}$ & ${0.400}_{-0.009}^{+0.009}$ (4.5\%) & ${0.400}_{-0.006}^{+0.005}$ (2.8\%) & ${0.400}_{-0.006}^{+0.005}$ (2.6\%) &  -  &  -  &  -  \\ 
D (Gpc) & ${20.01}_{-0.25}^{+0.24}$ (2.5\%) & ${20.00}_{-0.07}^{+0.07}$ (0.7\%) & ${20.00}_{-0.03}^{+0.03}$ (0.3\%) & ${19.98}_{-0.10}^{+0.09}$ (1.0\%) & ${20.00}_{-0.06}^{+0.06}$ (0.6\%) & ${20.00}_{-0.02}^{+0.02}$ (0.2\%) \\ 
$\iota$ (deg) & ${149.01}_{-0.82}^{+0.84}$ (1.12\%) & ${148.96}_{-0.16}^{+0.16}$ (0.22\%) & ${148.98}_{-0.16}^{+0.15}$ (0.21\%) & ${148.94}_{-0.47}^{+0.42}$ (0.60\%) & ${148.97}_{-0.11}^{+0.11}$ (0.15\%) & ${148.97}_{-0.11}^{+0.11}$ (0.14\%) \\ 
$\phi$ (deg) & ${229.2}_{-2.0}^{+2.0}$ (1.7\%) & ${229.2}_{-1.2}^{+1.1}$ (1.0\%) & ${229.2}_{-1.1}^{+1.0}$ (0.9\%) & ${229.0}_{-4.0}^{+4.5}$ (3.7\%) & ${229.2}_{-2.3}^{+2.3}$ (2.0\%) & ${229.0}_{-2.2}^{+2.1}$ (1.9\%) \\ 
$\beta$ (deg) & ${-34.4}_{-0.6}^{+0.7}$ (3.9\%) & ${-34.4}_{-0.4}^{+0.4}$ (2.1\%) & ${-34.4}_{-0.0}^{+0.0}$ (0.2\%) & ${-34.3}_{-0.4}^{+0.4}$ (2.1\%) & ${-34.4}_{-0.4}^{+0.4}$ (2.3\%) & ${-34.4}_{-0.0}^{+0.0}$ (0.2\%) \\ 
$\lambda$ (deg) & ${34.4}_{-0.9}^{+0.8}$ (5.1\%) & ${34.4}_{-0.5}^{+0.5}$ (2.7\%) & ${34.4}_{-0.6}^{+0.6}$ (3.5\%) & ${34.3}_{-0.5}^{+0.5}$ (2.9\%) & ${34.4}_{-0.5}^{+0.5}$ (3.1\%) & ${34.4}_{-0.7}^{+0.6}$ (3.6\%) \\ 
$\psi$ (deg) & ${22.9}_{-1.3}^{+1.1}$ (10.7\%) & ${22.9}_{-0.3}^{+0.3}$ (2.5\%) & ${22.9}_{-0.3}^{+0.3}$ (2.9\%) & ${23.0}_{-1.6}^{+1.6}$ (14.2\%) & ${22.9}_{-0.6}^{+0.6}$ (5.1\%) & ${22.9}_{-0.6}^{+0.6}$ (5.1\%) \\ 
\hline 
Sampling time & 6.1 h & 10.0 h & 9.5 h & 13.9 h & 1 d 8.3 h & 1 d 0.5 h \\ 
\# samples & 10080 & 10668 & 10108 & 18088 & 16436 & 10192 \\ 
\hline
\end{tabular}
\caption{Median values and error bars for each parameter, with the relative error between the median and the injected value indicated in brackets. The results are compared across runs with and without subdominant harmonics, with and without precessing spins, and with the inclusion of unequal-arms (UA).}
\label{tab:22_vs_HM}
\end{table*}

\begin{table*}[ht]
\setlength{\tabcolsep}{ 2pt}  
\renewcommand{\arraystretch}{1.4}  
\hspace*{-1cm}
\begin{tabular}{ l | r r r r r r }
\hline
Parameter & THM-A & THM-AE & THM-AET-G2 & TPHM-A & TPHM-AE & TPHM-AET-G2 \\
\hline
$m_1\: (10^6M_\odot)$ & ${1.0000}_{-0.0014}^{+0.0014}$ (0.28\%) & ${1.0000}_{-0.0008}^{+0.0007}$ (0.15\%) & ${1.0000}_{-0.0007}^{+0.0007}$ (0.15\%) & ${0.9997}_{-0.0021}^{+0.0018}$ (0.39\%) & ${0.9999}_{-0.0011}^{+0.0012}$ (0.23\%) & ${1.0000}_{-0.0009}^{+0.0010}$ (0.19\%) \\ 
$m_2\: (10^5M_\odot)$ & ${7.0002}_{-0.0101}^{+0.0097}$ (0.28\%) & ${7.0000}_{-0.0053}^{+0.0056}$ (0.16\%) & ${7.0000}_{-0.0051}^{+0.0050}$ (0.15\%) & ${7.0017}_{-0.0125}^{+0.0146}$ (0.39\%) & ${7.0004}_{-0.0082}^{+0.0074}$ (0.22\%) & ${7.0000}_{-0.0067}^{+0.0066}$ (0.19\%) \\ 
$a_1$ &  -  &  -  &  -  & ${0.701}_{-0.006}^{+0.006}$ (1.8\%) & ${0.700}_{-0.004}^{+0.004}$ (1.2\%) & ${0.700}_{-0.004}^{+0.005}$ (1.2\%) \\ 
$\theta_1$ (deg) &  -  &  -  &  -  & ${40.0}_{-0.6}^{+0.6}$ (2.9\%) & ${40.1}_{-0.3}^{+0.3}$ (1.7\%) & ${40.1}_{-0.4}^{+0.3}$ (1.8\%) \\ 
$\phi_1$ (deg) &  -  &  -  &  -  & ${38.1}_{-9.2}^{+8.2}$ (43.4\%) & ${38.6}_{-5.9}^{+6.2}$ (30.1\%) & ${38.9}_{-6.0}^{+6.7}$ (31.7\%) \\ 
$a_2$ &  -  &  -  &  -  & ${0.398}_{-0.017}^{+0.016}$ (8.2\%) & ${0.399}_{-0.011}^{+0.011}$ (5.5\%) & ${0.399}_{-0.012}^{+0.011}$ (5.9\%) \\ 
$\theta_2$ (deg) &  -  &  -  &  -  & ${22.9}_{-1.7}^{+1.9}$ (15.6\%) & ${22.7}_{-1.3}^{+2.0}$ (14.6\%) & ${22.8}_{-1.3}^{+2.1}$ (14.8\%) \\ 
$\phi_2$ (deg) &  -  &  -  &  -  & ${32.7}_{-22.5}^{+30.5}$ (231.2\%) & ${32.3}_{-20.7}^{+26.1}$ (204.2\%) & ${31.6}_{-21.0}^{+23.9}$ (196.2\%) \\ 
$s_{1z}$ & ${0.700}_{-0.005}^{+0.004}$ (1.3\%) & ${0.700}_{-0.003}^{+0.003}$ (0.9\%) & ${0.700}_{-0.003}^{+0.003}$ (0.8\%) &  -  &  -  &  -  \\ 
$s_{2z}$ & ${0.400}_{-0.009}^{+0.009}$ (4.3\%) & ${0.400}_{-0.006}^{+0.006}$ (2.8\%) & ${0.400}_{-0.005}^{+0.005}$ (2.7\%) &  -  &  -  &  -  \\ 
D (Gpc) & ${20.01}_{-0.14}^{+0.13}$ (1.3\%) & ${20.00}_{-0.08}^{+0.07}$ (0.7\%) & ${20.00}_{-0.07}^{+0.07}$ (0.7\%) & ${20.01}_{-0.16}^{+0.13}$ (1.5\%) & ${20.00}_{-0.07}^{+0.06}$ (0.7\%) & ${20.00}_{-0.06}^{+0.07}$ (0.6\%) \\ 
$\iota$ (deg) & ${148.97}_{-0.27}^{+0.26}$ (0.36\%) & ${148.97}_{-0.17}^{+0.16}$ (0.22\%) & ${148.97}_{-0.16}^{+0.16}$ (0.21\%) & ${148.96}_{-0.13}^{+0.16}$ (0.20\%) & ${148.97}_{-0.11}^{+0.11}$ (0.15\%) & ${148.97}_{-0.11}^{+0.11}$ (0.15\%) \\ 
$\phi$ (deg) & ${229.2}_{-1.9}^{+1.7}$ (1.6\%) & ${229.2}_{-1.1}^{+1.1}$ (1.0\%) & ${229.1}_{-1.1}^{+1.1}$ (0.9\%) & ${228.9}_{-3.5}^{+3.3}$ (3.0\%) & ${228.9}_{-2.2}^{+2.1}$ (1.9\%) & ${229.1}_{-2.6}^{+2.2}$ (2.1\%) \\ 
$\beta$ (deg) & ${-34.5}_{-0.9}^{+1.0}$ (5.7\%) & ${-34.4}_{-0.3}^{+0.4}$ (2.1\%) & ${-34.4}_{-0.4}^{+0.4}$ (2.1\%) & ${-34.5}_{-1.0}^{+1.2}$ (6.5\%) & ${-34.4}_{-0.4}^{+0.5}$ (2.5\%) & ${-34.3}_{-0.4}^{+0.4}$ (2.3\%) \\ 
$\lambda$ (deg) & ${34.5}_{-1.3}^{+1.3}$ (7.4\%) & ${34.4}_{-0.5}^{+0.5}$ (2.7\%) & ${34.4}_{-0.5}^{+0.5}$ (2.8\%) & ${34.5}_{-1.6}^{+1.4}$ (8.6\%) & ${34.4}_{-0.6}^{+0.6}$ (3.4\%) & ${34.3}_{-0.5}^{+0.6}$ (3.1\%) \\ 
$\psi$ (deg) & ${22.9}_{-0.6}^{+0.6}$ (5.0\%) & ${22.9}_{-0.3}^{+0.3}$ (2.4\%) & ${22.9}_{-0.3}^{+0.3}$ (2.4\%) & ${22.9}_{-1.0}^{+1.0}$ (9.0\%) & ${22.9}_{-0.7}^{+0.6}$ (5.4\%) & ${22.9}_{-0.6}^{+0.6}$ (5.3\%) \\ 
\hline 
Sampling time & 7.6 h & 8.0 h & 10.9 h & 1 d 9.1 h & 1 d 4.2 h & 1 d 17.8 h \\ 
\# samples & 10304 & 10052 & 10108 & 10136 & 14476 & 10136 \\ 
\hline
\end{tabular}
\caption{Median values and error bars for each parameter, with the relative error between the median and the injected value indicated in brackets. The results are compared across runs with channels A and AE and the second TDI generation (G2) for AET.}
\label{tab:tdis}
\end{table*}

\section{Conclusions} \label{sec:conclusions}

We have performed the first full Bayesian parameter estimation study for LISA MBHBs employing a time domain response and including the effects of subdominant harmonics and precession. The time domain approach offers a more robust theoretical formalism compared to the approximated \textit{transfer functions} formalism used in the Fourier domain. It facilitates a straightforward inclusion of various physical effects in the waveform, such as precession, eccentricity or memory, and provides a basis to test and validate the accuracy of the Fourier domain formalism.

Until now, time domain methods have been computationally prohibitive for LISA parameter estimation. We have significantly reduced the computational cost of waveform generation and likelihood evaluation to make this approach feasible. This was achieved using a new Python implementation of the \textsc{IMRPhenomT} waveform family with GPU support, which parallelizes the computation of each waveform’s time array. Additionally, the precession module in this waveform model now applies a more efficient algorithm to compute the polarizations, distinct from that in the LALSuite implementation used by the LVK. The LISA arm response and TDI channels are computed directly in the time domain also through a GPU-accelerated version of the LISA Data Challenge tools. Both the Fourier transforms and the inner product calculations required for the likelihood evaluation are processed on the GPU, with the final likelihood value passed to the Bayesian inference package \textsc{Bilby} for stochastic sampling and likelihood maximization across the parameter space.

We utilized the \textsc{ptemcee} sampler with Fisher initialization and broad uniform priors to conduct zero-noise injections both for aligned-spin and precessing systems. Our analysis investigates the influence of subdominant harmonics and unequal-arm orbits, as well as different TDI channels and TDI generations on parameter recovery. To further validate these findings, we compared the resulting Bayesian posteriors with those derived from the standard Fisher analysis, observing overall consistency with the expected theoretical behaviour across all parameters.

Our results demonstrate LISA’s ability to measure individual mass components with a precision of 0.1\% and spin components to within 1\%, excluding the azimuthal spin angles and tilt angle of the secondary spin. The reduced accuracy in spin angles is likely attributable to the single-spin approximation used for the Euler angles in the precessing model. We also observed that including subdominant harmonics effectively resolves the degeneracy between inclination and distance, particularly in the aligned-spin case, enhancing the accuracy of parameter estimation across all measured parameters for both aligned and precessing spins systems.

This work represents the first Bayesian inference study to explicitly assess the impact of unequal-arms, different TDI channels and TDI generations on LISA parameter estimation for MBHBs. Results indicate that the realistic orbits with unequal-arms improve the measurement of the sky localization and distance. We also observe that the T channel does not influence posterior distributions, which aligns with expectations, as this channel is designed to be noise-dominated, and our injected signals are noise-free. Similarly, no significant differences emerged between TDI generations in the current setup. However, this does not rule out the need for further development of TDI channels and generations, since the influence of TDI configurations is likely to be much more pronounced in analyses involving realistic noise. Our next objective is to extend this study by incorporating Gaussian noise into the signal, which remains unexplored even in Fourier domain approaches.


Our analysis was executed on a single GPU, but future work will involve running multiple Markov chains Monte Carlo in parallel across several GPUs, which will enhance computational efficiency. Additionally, we aim to introduce GPU support for alternative precessing prescriptions within the \textsc{IMRPhenomT} model, potentially improving the accuracy of certain parameter measurements. Current development also includes incorporating eccentricity and the memory effect into our model, which our parameter estimation pipeline can accommodate seamlessly. This enhancement would enable the first studies of these effects on LISA parameter estimation, broadening our capacity to gain richer physical information from the signal and advancing our understanding of the underlying astrophysical processes.

We recognize that optimizing sampler settings could further reduce sampling time and improve posterior convergence; however, this will be addressed in future studies, alongside trials of alternative samplers to evaluate their effectiveness.
One current limitation is the necessity of initializing the sampler with Fisher estimates to enhance convergence speed. This requirement could potentially be addressed by further reducing the computational cost associated with likelihood evaluations. In this regard, we are exploring more efficient algorithms for waveform and likelihood computation, including nonuniform time grids and interpolation methods such as Multibanding, Relative Binning or \textsc{FANTA}\footnote{\textsc{FANTA} (Fourier Analytical Transform of a piecewise Approximant) leverages the Fourier transform’s linearity to analytically integrate a simplified form of the waveform function across each time interval.}. By implementing these acceleration techniques, we also aim to extend this framework to lower-mass systems with longer-duration signals, such as those from stellar-mass black holes.

The next steps and challenges can be summarized as
\begin{itemize}
    \item Perform injections in fully realistic LISA noise.
    \item So far the injected signal and recovery have been performed with the same waveform model and LISA settings. The next step is to use different settings for the recovery.
    \item Optimization of sampler settings for general sampler initialization.
    \item Parallelization of sampling over multiple GPUs.
    \item Further speed-up of waveform and likelihood evaluation with nonuniform grids algorithms.
\end{itemize}

\section*{Data Availability}
Data supporting the findings of this article are openly available \cite{data}.

\section*{Acknowledgements} \label{sec:acknowledgements}
We thank Jorge Valencia for finding a bug in the LISA response code and for useful discussions. CG is supported by the Swiss National Science Foundation (SNSF) Ambizione grant PZ00P2\_223711 and was also supported by Sinergia grant 213497. ST is supported by the Swiss National Science Foundation Ambizione Grant Number : PZ00P2-202204.  We gratefully acknowledge support from the CNRS/IN2P3 Computing Center (Lyon - France) for providing computing and data-processing resources needed for this work.

\bibliography{ref}

\end{document}